\pdfoutput=1 \pdfsuppresswarningpagegroup=1
\documentclass[
 aps,%
 prb,
 preprint,% >1 column
 reprint,% >2 column
 groupedaddress,% 
 superscriptaddress,%
 showpacs,%
 letterpaper,%
 footinbib,%
 floatfix,%
 noeprint%
]{revtex4-1}
\RequirePackage{amsmath,amsfonts,amssymb,bm,braket}
\RequirePackage{graphicx}
\RequirePackage{hyperref}
\hypersetup{%
  breaklinks = {true},
  citecolor = {blue},
  colorlinks = {true},
  linkcolor = {red},
  pdfcreator = {\LaTeX\ and \flqq hyperref\frqq},
}

\renewcommand{\vec}[1]{\text{$\boldsymbol{#1}$}}

\newcommand{\Fref}[1]{Fig.~\ref{#1}}
\newcommand{\Fsref}[1]{Figs.~\ref{#1}}
\renewcommand{\eqref}[1]{(\ref{#1})}
\newcommand{\Eqref}[1]{Eq.~(\ref{#1})}

\renewcommand{\Ref}{Ref.}
\newcommand{\Refs}{Refs.}
\newcommand{\Sref}[1]{Section~\ref{#1}}
\newcommand{\eg}{\emph{e.g.}{}}
\newcommand{\ie}{\emph{i.e.}{}}
\newcommand{\etal}{\emph{et al.}}

\newcommand{\bk}{\vec{k}}
\newcommand{\bq}{\vec{q}}
\newcommand{\br}{\vec{r}}
\newcommand{\bG}{\vec{G}}
\newcommand{\bt}{\vec{t}}
\renewcommand{\Re}{\operatorname{Re}}
\renewcommand{\Im}{\operatorname{Im}}

\begin{document}

\title{Expeditious computation of nonlinear optical properties of arbitrary 
order with native electronic interactions in the time domain}

\author{Emilia Ridolfi}
\affiliation{%
  Centre for Advanced 2D Materials, National University of Singapore,
  6 Science Drive 2, Singapore 117546, Singapore}
\affiliation{%
  BGeosys, Department of Geoscience, Environment \& Society (DGES), 
  Universit\'e Libre de Bruxelles, 50 Avenue F.D. Roosevelt, Brussels 1050, 
  Belgium}
  
\author{Paolo E. Trevisanutto}
\affiliation{%
  Centre for Advanced 2D Materials, National University of Singapore,
  6 Science Drive 2, Singapore 117546, Singapore}
\affiliation{%
  European Centre for Theoretical Studies in Nuclear Physics and 
  Related Areas (ECT*-FBK)}
\affiliation{%
  Center for Information Technology, Bruno Kessler Foundation, 
  Trento 38123, Italy}

\author{Vitor~M.~Pereira}
\email[Corresponding author: ]{vpereira@nus.edu.sg}

\affiliation{%
  Centre for Advanced 2D Materials, National University of Singapore,
  6 Science Drive 2, Singapore 117546, Singapore}
\affiliation{%
  Department of Physics, National University of Singapore,
  2 Science Drive 3, Singapore 117542, Singapore}

\date{\today}

\begin{abstract}
We adapted a recently proposed framework to characterize the optical response of 
interacting electrons in solids in order to expedite its computation without 
compromise in accuracy at the microscopic level. 
Our formulation is based on reliable parameterizations of Hamiltonians and 
Coulomb interactions, which allows economy and flexibility in obtaining response 
functions. It is suited to computing the optical response to fields of 
arbitrary temporal shape and strength, to arbitrary order in the field, and 
natively accounts for excitonic effects.
We demonstrate the approach by computing the frequency-dependent 
susceptibilities of MoS$_2$ and hexagonal BN monolayers up to the 
third-harmonic.
Grounded on a generic non-equilibrium many-body perturbation theory, this 
framework allows extensions to handle generic interaction models or to 
describe electronic processes taking place at ultrafast time scales.
\end{abstract}

\maketitle

%-------------------------------------------------------------------------------
\section{Introduction}
%-------------------------------------------------------------------------------

The nonlinear optical response of crystalline materials is a rich and attractive 
playground for optoelectronic applications \cite{Boyd2008, Luppi2016}. 
Traditionally, only a few select bulk crystals were considered of practical 
interest because nonlinearities tend to be weak effects \cite{Boyd2008}. But the 
recent advent and proliferation of countless two-dimensional (2D) materials 
have enormously broadened the range of platforms for studying nonlinear optical 
properties \cite{Wang2012d, Xia2014b, Luppi2016, WangReview2018}. Several 
classes of 2D crystals host a range of new electronic effects and 
functionalities, including nontrivial topological characteristics \cite{Xu2014, 
Schaibley2016} and enhanced electronic interactions, which arise from their 
strict two-dimensionality \cite{Cudazzo2011, Thygesen2017, Kotov2012}. One 
specific implication of enhanced interactions is that excitons become an 
essential element of the optical response of 2D semiconductors over extended 
frequency ranges \cite{WangReview2018, Ridolfi2018}. Moreover, excitons play a  
particularly critical role in high-order effects such as harmonic generation.
For example, 2D transition metal dichalchogenide (TMD) semiconductors harbour 
bound excitons at energies practically resonant with lasers in standard use, 
thereby showing a consistently strong and easy-to-access nonlinear response 
\cite{WangReview2018, Liu2016b}. For these reasons, the ability to theoretically 
understand and model these nonlinear characteristics is of very high current 
interest. 

The strong optical response of 2D semiconductors also means they can more 
efficiently be driven out of equilibrium by optical excitation. This is of 
fundamental interest to probe and understand microscopic mechanisms underlying 
a number of proposed functionalities, including their valley and spin 
relaxation characteristics that are crucial for applications in valleytronics 
and spintronics, respectively \cite{Xu2014, Schaibley2016}. Ultrafast 
spectroscopy experiments are a versatile and proven tool in this regard 
\cite{Huber2001, Kampfrath2013, Kim2014b, Hao2016, Ye2016b} which, in turn, 
demands realistic and accurate theoretical methods to model the microscopic 
transient response of electrons on fast timescales. Here, too, interactions are 
essential, not only to capture the correct excitations, but also the hot 
relaxation pathways at short timescales. 

Unfortunately, handling interactions within accurate descriptions of the 
underlying electronic structure, such as in \emph{ab initio} density 
functional theory (DFT) methods, is a perennially challenging problem, both 
methodologically and computationally \cite{Onida2002}. 
Explicitly accounting for electronic interactions in a systematic perturbative 
expansion with respect to an external field quickly becomes a cumbersome task 
due to the combinatorial proliferation of terms involving both matrix elements 
and excitonic wavefunctions at each order \cite{Chang2001, Leitsmann2005}. If 
the coupling to the external field is described in the length-gauge, that 
proliferation is even more severe due to the need to explicitly separate intra- 
and inter-band transitions \cite{Sipe1995, Pedersen2015hBN}. The typical 
development of the perturbative series in the frequency domain will also be 
inadequate to describe phenomena that are intrinsically of the \emph{time 
domain}, such as transient processes in response to intense fields. In addition, 
strong fields can drive the electronic system out of equilibrium which thus 
requires a nonequilibrium theoretical framework. Finally, for a direct 
connection with time-resolved spectroscopy, it is desirable to develop 
capabilities to describe a system's reaction not only to monochromatic 
continuous-wave excitation but to arbitrary time-dependent fields.

In view of these challenges, the current paper demonstrates a good compromise, 
between the microscopic accuracy of DFT-based electronic structure calculations 
and numerical expediency when modeling the response of an electronic system to 
strong electromagnetic fields.
There have been recent developments to approach the net electric response in the 
time domain \emph{ab initio}. Such approaches benefit from the unbiased 
nature of DFT calculations \cite{Onida2002} as well as their accuracy when 
extended with quasiparticle corrections within many-body perturbation theory 
\cite{Souza2004, Takimoto2007, Ding2013, Pal2011, Perfetto2015, Attaccalite2011, 
Attaccalite2013, Attaccalite2014, Attaccalite2016, Tancogne2016, Luppi2016}. 
While these represent remarkable conceptual and pragmatic progress, practical 
implementations remain arduous because of the complexity inherent to a 
self-consistent description of (i) many-body interactions, (ii) deviations from 
equilibrium and (iii) relaxation processes. The enhanced Coulomb interaction in 
2D materials is a further challenge due to more stringent convergence demands 
\cite{Leitsmann2005, Huser2013, Qiu2013, Qiu2015-erratum}.  
This effort will benefit from implementations that can deliver faster results 
with the same level of accuracy as a fully \emph{ab initio} approach.

This paper contributes in that direction. It hinges on the framework originally 
proposed by Attaccalite \etal{}, which combines DFT and nonequilibrium 
many-body perturbation theory to compute the time-dependent polarization in 
electronic systems excited by arbitrary electric fields \cite{Attaccalite2011}. 
Our approach, however, trades the self-consistent calculation of the Kohn-Sham 
Hamiltonian, the dynamically screened Coulomb interaction and $GW$ quasiparticle 
corrections, by parameterized tight-binding (TB) and screening models 
demonstrated to be reliable and accurate in the context of 2D 
materials\,---\,especially in the treatment of the excitonic degrees of freedom 
\cite{Ridolfi2018}. 
The key features of our approach are: the ability to compute, 
non-perturbatively, the response to fields with arbitrary strength and temporal 
profile; the native inclusion of electronic interactions in the time evolution 
of the excited states; and the explicit consideration that the external field 
drives the electronic system out-of-equilibrium during its time evolution. In 
weak external fields, it becomes equivalent to a perturbative response 
calculation based on the solution of the Bethe-Salpeter equation (BSE) 
\cite{Attaccalite2011}. Its power, though, is best revealed in the ability to 
extract nonlinear susceptibilities to arbitrary order in a one-shot computation 
with no more technical effort than what is necessary to obtain the linear 
response.

Formulated in terms of Green's functions (GFs) and having all the effects of 
interactions and relaxation encoded into an electronic self-energy, this 
strategy lends itself to systematic extensions beyond the presently explored 
approximations. But, above all, the fact it hinges on parameterized\,---\,yet 
accurate\,---\,Hamiltonians makes this not only an expedite but also a flexible 
framework to tackle the theoretical description of strong nonlinearities. A case 
in point would be when such calculations need to cover a large parameter space 
of interest for a given material (\eg, as a function of strain or doping). 
Another typical use-case is large-scale deployment: Several catalogs currently 
in development by different materials database projects \cite{Jain2013, 
Saal2013, Rasmussen2015, Mounet2016, Zhou2019a} lack linear and nonlinear 
optical response functions. Whereas a fully \emph{ab initio} approach would be 
computationally prohibitive in both of these scenarios, an implementation as we 
describe below is well within reach of current computational capabilities.

We demonstrate the concept with an application to monolayers of molybdenum 
disulfide (MoS$_2$) and hexagonal boron nitride (hBN). The former is 
representative of the important family of TMD semiconductors, which we chose to 
explicitly illustrate that excellent agreement is possible. Moreover, as a 
reasonable bandstructure parameterization to study harmonic generation across 
this family frequently requires consideration of 8--10 bands \cite{Pedersen2014, 
Wu2015, Ridolfi2018}, it further illustrates the expediency of this approach 
with a relatively demanding model parameterization. For hBN, we chose a minimal 
2-band TB model to establish the importance of a truly non-equilibrium 
formulation where, in addition to intra-band matrix elements of the dipole 
operator \cite{Pedersen2015hBN}, the time-evolution of the electronic 
populations should be explicitly accounted for in calculations of nonlinear 
optical properties using minimal TB parameterizations.

The remainder of this paper is organized as follows. For conceptual 
self-consistency, \Sref{sec:methodology} revisits the key aspects of the 
methodology first proposed in \Ref~\onlinecite{Attaccalite2011} as a 
time-domain version of the BSE. In addition to outlining the development of the 
equation of motion for the relevant GFs, its subsections describe our specific 
approach to the TB parameterizations within the Slater-Koster scheme 
\cite{Slater1954}, how the relevant matrix elements are computed in this 
framework, the parameterization of the screened Coulomb interaction, and the 
approximation to the effective self-energy. For pedagogical purposes, we include 
additional subsections covering practical aspects of the Fourier analysis, of 
two phenomenological relaxation schemes, as well as additional notes related to 
our implementation.
\Sref{sec:results} contains the core of our results by applying the technique to 
the cases of MoS$_2$ and hBN. Its subsections describe the specific 
parameterizations used for each material, a brief overview of the main spectral 
features in their optical response, an application of the technique in a 
one-shot scenario to demonstrate it recovers the absorption spectrum 
alternatively calculated in a Kubo formalism from the solution of the BSE, and a 
demonstration of its application to extract high-harmonic susceptibilities.
Prior to our conclusion in \Sref{sec:conclusion}, \Sref{sec:discussion} 
addresses the main trade-offs of time-domain calculations, their intrinsic 
adequacy to probe ultrafast electronic processes, the role of intra-band matrix 
elements, the need to employ a nonequilibrium distribution, and the method's 
overall numerical scaling.

% ------------------------------------------------------------------------------
% FIGURE BEGINS
% ------------------------------------------------------------------------------
\begin{figure*}
\centering
\includegraphics[width=0.9\textwidth]{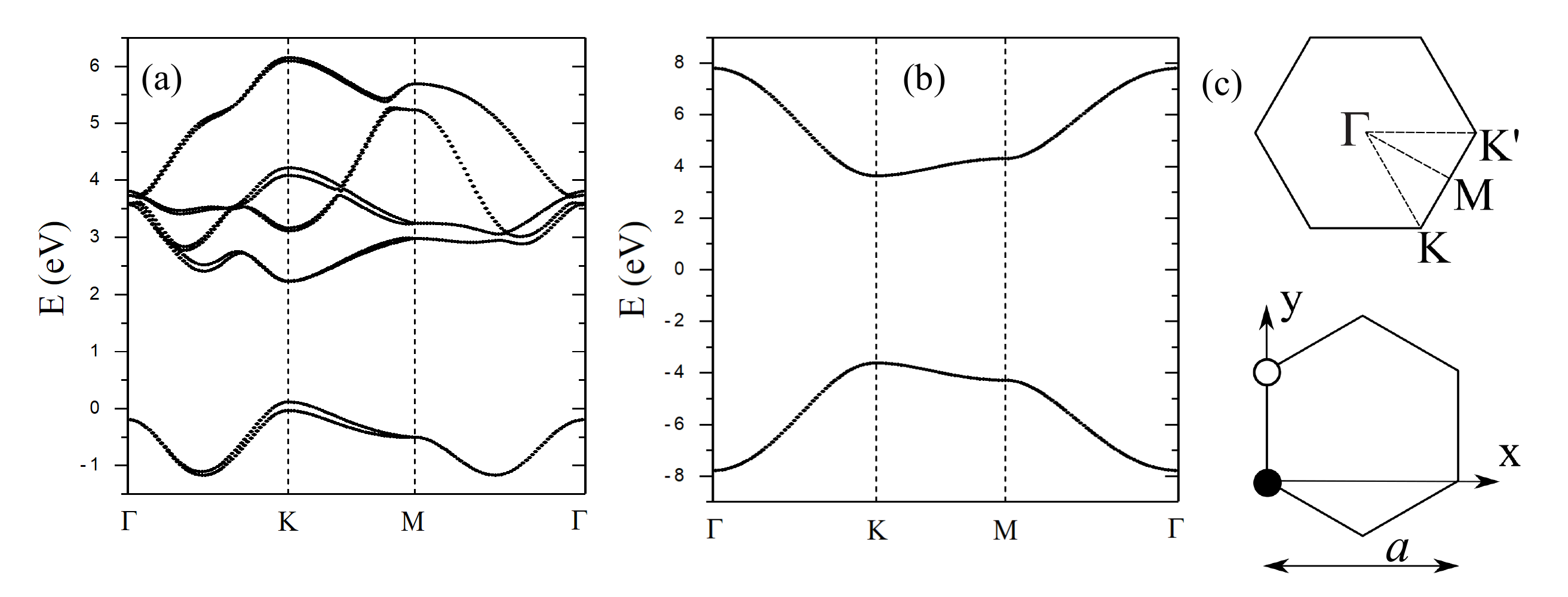} 
\caption{
(a) TB-derived band structure (including SO coupling) of the MoS$_{2}$ 
monolayer according to the Slater-Koster scheme discussed in the text and 
reported earlier in \Ref~\onlinecite{Ridolfi2015}. (The Fermi energy is set to 
0.)
(b) TB-derived band structure of hBN (spin-degenerate) according to the 
parameterization discussed in the text and reported in 
\Ref~\onlinecite{Galvani2016}. (The Fermi energy is set to 0.)
(c) Top: Illustration of the Brillouin zone for both MoS$_2$ and hBN. 
Bottom: Representation of our orientation of the honeycomb lattice in real 
space with the positions of the $A$ and $B$ sites highlighted by the filled and 
empty circles. While $A$ and $B$ represent the positions of N and B atoms in 
hBN, in the case of MoS$_2$, we can associate them with a top view of the Mo 
and S atoms, respectively.
}
\label{fig:bands}
\end{figure*}
% ------------------------------------------------------------------------------

%-------------------------------------------------------------------------------
\section{Methodology} 
\label{sec:methodology}
%-------------------------------------------------------------------------------

The central goal is to compute the time-dependent polarization induced in a 
non-polar crystal, $\bm{P}(t)$ (dipole moment per unit area of the crystal), in 
response to a time-dependent electromagnetic field, $\bm{E}(t)$. The field may
have generic magnitude and time dependence (\ie, neither small nor sinusoidal), 
and we wish the computed $\bm{P}(t)$ to include the major effects of 
electron-electron interactions\,---\,fundamental for an adequate 
description of the optical properties of 2D semiconductors, where excitonic 
renormalization effects are very pronounced and dominate the optical spectra 
\cite{Thygesen2017, WangReview2018}. 
In the most general terms, the response to a field of arbitrary strength is 
determined by the $n$-th order susceptibilities according to 
\begin{align}
  \epsilon_0^{-1} P^\lambda(t) & = \int^t \!\! dt_1 \,
    \chi^{(1)}_{\lambda\alpha}(t,t_1) E^{\alpha}(t_1)  
  \nonumber \\
  & + \int^t \!\! dt_1 \! \int^{t_1} \!\!\! dt_2 \,
    \chi^{(2)}_{\lambda\alpha\beta}(t,t_1,t_2) E^{\alpha}(t_1) 
    E^{\beta}(t_2)
  \nonumber \\
  & + ... ,
  \label{eq:P-vs-chi-t}
\end{align}
where $\epsilon_{0}$ is the vacuum dielectric permittivity and $\lambda$, 
$\alpha$, $\beta$ are Cartesian components. Being non-perturbative, the 
calculated $\bm{P}(t)$ will contain the information of all 
nonlinear susceptibilities $\chi^{(n)}$ (which can be extracted by appropriate 
post-processing) with excitonic effects included from the outset. To capture 
microscopic coherence in $\bm{P}(t)$ to all orders in the external field 
requires a nonequilibrium framework \cite{Schmitt-Rink1988}.

Our approach is strongly inspired by the technique and approximations proposed 
by Attaccalite \etal{} for a fully \emph{ab initio} strategy to obtain the 
optical response in the time domain based on nonequilibrium GFs 
\cite{Attaccalite2011}. However, we depart from their original formulation by 
(i) considering a TB parameterization of the quasi-particle-renormalized 
Hamiltonian, as well as by (ii) describing the Coulomb interactions by a 
parameterized screened potential. This expedites the numerical time-integration 
and makes calculations at each time step less memory-demanding, especially if 
the desired energy range comprises many bands and/or a large number of $\bk$ 
points in the sampling of the Brillouin zone (BZ)\footnote{That is frequently 
the case for 2D semiconductors; see for example \Refs~\onlinecite{Huser2013, 
Qiu2013, Qiu2015-erratum}.}. Our objective in the present paper is to 
explicitly show this approach to be both reliable and efficient.

Throughout this article we consider that the target system remains spatially  
homogeneous under the influence of the external radiation field. This is 
appropriate since the frequencies of interest are of the order of the optical 
bandgap (typically in the infrared-to-visible range), with associated 
wavelengths much larger than the atomic distances. As we concentrate on strictly 
2D crystals, all our vector quantities are restricted to the $xy$ plane, which 
coincides with the crystalline sheet.

The polarization can be expressed in terms of the one-particle reduced density 
matrix as \cite{Resta1994}
\begin{equation}
  \bm{P}(t) = e \int \!\br \hat{\rho}(\br) d\br
    = \frac{e}{A} \sum_{mn\bk} \br_{mn\bk} \, \rho_{mn\bk}(t),
  \label{eq:P-def-rho}
\end{equation}
where $e<0$ is the charge of the electron; $A$ is the area of the crystal 
($A=A_c N_k^2$ for $N_k^2$ unit cells and area per cell $A_c\equiv 
\sqrt{3}a^2/2$); $\rho_{mn\bk}(t) \equiv \langle a^\dagger_{m\bk}(t) 
a_{n\bk}(t)\rangle$ is the reduced density matrix; the Heisenberg operator 
$a^\dagger_{m\bk}(t)$ creates an electron in the Bloch eigenstate 
$\psi_{m\bk}(\br)\equiv\braket{\br|m\bk}$ at time $t$; $\br_{mn\bk} \equiv 
\bra{m\bk}\hat{\bm{r}}\ket{n\bk}$. As we are interested in a non-equilibrium 
description, we introduce the two-time \emph{lesser} GF 
\cite{Kadanoff1962,Kremp2005} in the Bloch representation:
\begin{multline}
  G^{<}_{mn\bk}(t,t') \equiv i\langle a^\dagger_{n\bk}(t') a_{m\bk}(t) \rangle
  \\
  = i \int \!\! d\br d\br' \, \psi^*_{m\bk}(\br) \psi_{n\bk}(\br') \, 
  \langle \hat{\psi}^\dagger(\br',t') \hat{\psi}(\br,t) \rangle .
  \label{eq:Gless-def} 
\end{multline}
Here, $\hat{\psi}(\br,t)$ represents the electronic field operator and 
$i\, \langle \hat{\psi}^\dagger(\br',t') \hat{\psi}(\br,t) \rangle 
{\,\equiv\,} G^<(\br t,\br' t')$, which is the real-space representation of 
the lesser GF. Note that the time-diagonal component of $G^{<}_{mn\bk}(t,t')$ 
coincides with the reduced density matrix, $\hat{\rho}(t)$:
\begin{equation}
  G^{<}_{mn\bk}(t) \equiv \lim_{t'\to t^+}G^{<}_{mn\bk}(t,t') 
  = i \rho_{nm\bk}(t).
\end{equation}
Hence, \Eqref{eq:P-def-rho} can be recast as
\begin{equation}
  \bm{P}(t) = -\frac{ie}{A} \sum_{mn\bk} \br_{mn\bk} \, G^{<}_{nm\bk}(t).
  \label{eq:P-def-G}
\end{equation}
The central problem is thus determining the time dependence of 
$G^{<}_{nm\bk}(t)$ under the influence of an external field of arbitrary 
strength and arbitrary temporal profile.

%-------------------------------------------------------------------------------
\subsection{Equation of motion for the distribution function}
%-------------------------------------------------------------------------------

Many-body electronic excitations in response to a time-dependent external field 
are most systematically handled with GF techniques. To obtain $\bm{P}(t)$ in a 
non-perturbative way implies that our description must properly handle 
arbitrarily strong fields (and, of course, in experiments, probing the nonlinear 
susceptibilities does require intense laser fields \cite{Boyd2008}). 
Strong fields are bound to drive the statistical system out of thermodynamic 
equilibrium. An accurate description of the coherent microscopic processes 
therefore requires a nonequilibrium GF formalism \cite{Schmitt-Rink1988}, which 
has been pioneered by Kadanoff and Baym \cite{Kadanoff1962}, and by Keldysh 
\cite{Keldysh1964a}.

Since details related to the derivation of the equation of motion for 
$G^{<}_{nm\bk}(t)$ have been discussed, for example, in 
\Refs~\onlinecite{Attaccalite2011} or \onlinecite{Schmitt-Rink1988}, we provide 
only a qualitative overview of its key aspects and assumptions. In addition to 
establishing the context for our numerical calculations in a conceptually 
self-contained way, this allows us to highlight the necessary 
adaptations necessary for our approach, which is based on parameterized 
Hamiltonians and interactions. 

Kadanoff and Baym provided a closed set of coupled equations for the time 
evolution of the different nonequilibrium GFs in terms of self-energies defined 
on distinct portions of the Keldysh contour; for practical calculations, the 
self-energies must be specified within an approximation scheme 
\cite{Kadanoff1962, Kremp2005}. A major simplifying step occurs by 
approximating $\Sigma^< = 0$ and $\Sigma^r = \Sigma^a \equiv \Sigma$, 
similarly to what happens in a collisionless and instantaneous scenario like 
Hartree-Fock \cite{Schmitt-Rink1988}. The equation for $G^{<}_{nm\bk}(t)$ 
then decouples and reads \cite{Attaccalite2011, Schmitt-Rink1988}: 
\begin{equation}
  i\hbar\frac{\partial}{\partial t}\bm{G}_{\bk}^{<}(t)
  = \Bigl[\bm{h}_{\bk}+\bm{U}_{\bk}(t)+\bm{\Sigma}_{\bk}[\bm{G}_{\bk}^{<}(t)], 
  \bm{G}_{\bk}^{<}(t) \Bigr],
  \label{eq:dGdt-1}
\end{equation}
where $[\bm{A},\bm{B}] \equiv \bm{A}\bm{B} - \bm{B}\bm{A}$. For notational 
simplicity, we employ bold symbols to denote matrices in the band indices: for 
example, $G_{mn\bk}^{<}(t)=[\bm{G}_{\bk}^{<}(t)]_{mn}$, and $h_{mn\bk} = 
[\bm{h_\bk}]_{mn} = \bra{m\bk}\hat{h}\ket{n\bk} = \delta_{mn} E_{m\bk}$, where 
$E_{m\bk}$ are the Bloch bands. 

The electronic system is defined here by the non-interacting Bloch Hamiltonian 
$\hat{h}$, and $\hat{U}(t)$ is the explicitly time-dependent external field. The 
total non-interacting Hamiltonian is thus 
\begin{equation}
  \hat{H}(t) = \hat{h} + \hat{U}(t).
  \label{eq:H}
\end{equation}
\Eqref{eq:dGdt-1} extends the most commonly employed framework known as 
``semiconductor Bloch equations'' (the dynamical equations for the reduced 
density matrix) \cite{Schafer2002, Sipe1995} with the addition of a self-energy 
$\hat{\Sigma}$, which can be non-Hermitian to capture relaxation processes. 
While the semiconductor Bloch equations are an equation-of-motion approach to 
the time evolution of $\hat{\rho}$ \cite{Schafer2002}, our formulation in terms 
of GFs and self-energies is best suited (through the machinery or GFs and 
diagramatics) for systematic study of different interaction and relaxation 
mechanisms without additional formal effort. The self-energy encodes all the 
many-body correlation and decoherence (in the imaginary part) effects; as these 
depend on the electronic occupations, $\hat{\Sigma}$ is a functional of 
$G_{mn\bk}^{<}(t)$ which is represented by the term
$\bm{\Sigma}_{\bk}[\bm{G}_{\bk}^{<}(t)]$ in \Eqref{eq:dGdt-1}.

We note that the Hamiltonian $\hat{h}$ is meant to be described in terms of a TB 
parameterization; but having it reflect the strictly non-interacting Bloch 
Hamiltonian is not optimal, for two main reasons. On the one hand, irrespective 
of whether the TB Hamiltonian is obtained from a DFT calculation or constrained 
directly by experiments, it will already incorporate electron-electron 
interactions (in the DFT case, the TB parameterization reflects at least 
the Kohn-Sham Hamiltonian, which already incorporates interactions at a basic 
level). On the other hand, it is desirable that $\hat{h}$ provides an accurate 
description of the ground state; in the case of a semiconductor\,---\,and 
particularly so for 2D materials\,---\,that requires incorporating the 
interaction-driven corrections to the quasiparticle dispersion beyond DFT 
\cite{Hybertsen1986}. Therefore, we take $\hat{h}$ and its spectrum, $E_{m\bk}$, 
to represent a TB parameterization of the ground state band structure which 
\emph{already} takes into account such corrections\,---\,for example, at the 
level of the $GW$ approximation \cite{Hedin1970,Hybertsen1986}, 
or as provided by hybrid functional approaches to DFT \cite{Becke1993}.

These considerations require a corresponding and consistent reassessment of the 
self-energy term in \Eqref{eq:dGdt-1} to avoid double-counting of interactions. 
We hence rewrite that equation as
\begin{equation}
  i\hbar\frac{\partial}{\partial t}\bm{G}_{\bk}^{<}(t)
  = \Bigl[ \bm{H}_{\bk}(t) + \bm{\Sigma}_{\bk}[G^{<}(t)]
    - \bm{\Sigma}_{\bk}[\tilde{G}^{<}], \,\bm{G}_{\bk}^{<}(t) \Bigr],
  \label{eq:dGdt-2}
\end{equation}
where $\tilde{G}^{<}$ represents the GF of the unperturbed system at 
equilibrium, 
\begin{equation}
  \tilde{G}^{<}_{mn\bk} \equiv G^{<}_{mn\bk}(t=0) 
    = i \delta_{mn} f_{m\bk},
  \label{eq:G-tilde}
\end{equation}
and $f_{m\bk}$ is the Fermi-Dirac distribution at energy $E_{m\bk}$. The effect 
of the term $- \bm{\Sigma}_{\bk}[\tilde{G}^{<}]$ is to subtract from the 
self-energy the quasiparticle corrections of the unperturbed system at 
equilibrium (which, according to the above discussion, should be already 
included in the TB parameterization for $\hat{h}$). In this way, the 
self-energy terms describe \emph{only} the correlation changes induced by the 
external field.
Equation \eqref{eq:dGdt-2} is our counterpart of Eq.~(11) proposed by 
Attaccalite \etal{} in \Ref~\onlinecite{Attaccalite2011}.

Note that the temperature only appears in the time evolution implicitly, via 
the Fermi-Dirac distribution that defines the initial condition 
\eqref{eq:G-tilde}. Zero and finite temperature calculations are thus on equal 
footing. Despite this, in the current work we set $T=0$ since we will benchmark 
our results against other zero-temperature calculations.

%-------------------------------------------------------------------------------
\subsection{Tight-binding parameterizations}
%-------------------------------------------------------------------------------

We rely on orthogonal TB Hamiltonians in the Slater-Koster formulation 
\cite{Slater1954} to represent $\hat{h}$, where the Bloch eigenstates states, 
$\psi_{n\bk}(\br)$, are expanded in terms of effective local atomic orbitals, 
$\phi_\alpha(\br)$, as follows:
\begin{align}
  \psi_{n\bk}(\vec{r}) & = \sum_{\alpha}C_{\alpha\bk}^{n} \, 
    \chi_{\alpha\bk}(\br),
  \nonumber \\
  \chi_{\alpha\bk}(\br) & \equiv \frac{1}{\sqrt{N_c}}
    \sum_{\vec{R}}e^{i(\bk\cdot\vec{R}+\theta_{\alpha \bk})} 
    \phi_{\alpha}(\vec{r}-\vec{R}-\vec{t}_\alpha).
  \label{eq:bloch}
\end{align}
The lattice vector $\vec{R}$ runs over all $N_c$ unit cells of the crystal, $n$ 
is the band index, $\alpha$ labels different orbitals within the unit cell which 
are centered at position $\vec{t}_\alpha$ relative to the cell's origin. Both 
$n$ and $\alpha$ run over the interval $[1,N]$, where $N$ is the dimension of 
the orbital basis considered. Although the phase factor $\theta_{\alpha \bk}$ 
can be fixed arbitrarily (for example $\theta_{\alpha\bk}=0$), one has to 
consistently carry that choice to the matrix elements of the dipole operator and 
screened Coulomb interaction (to be discussed below). For convenience we set 
$\theta_{\alpha\bk} = \bk\cdot\vec{t}_\alpha$ everywhere in this work 
\cite{Ventura2019}. 

The specific TB parameterizations used in our calculations for MoS$_2$ and BN 
will be discussed further below. The underlying bandstructures are shown in 
\Fref{fig:bands}.

%-------------------------------------------------------------------------------
\subsection{External field}
\label{sec:field}
%-------------------------------------------------------------------------------

We express the perturbation due to the external radiation field, $\bm{E}(t)$, 
in the dipole approximation and length gauge \cite{Sipe1995}:
\begin{equation}
  \hat{U}(t) \equiv -e\, \hat{\br}\cdot\bm{E}(t).
  \label{eq:U-field}
\end{equation}
Its matrix elements in the Bloch basis are $U_{mn\bk}(t) = 
-e\,\br_{mn\bk}\cdot\bm{E}(t)$ and thus require the computation of 
$\br_{mn\bk}\equiv\bra{m\bk}\hat{\br}\ket{n\bk}$. We shall consider only 
inter-band transitions and neglect all intra-band matrix elements, $\br_{mm\bk}$ 
(this point is revisited later). In that case, from the definition of the 
velocity operator $i\hbar\hat{\bm{v}} = [\hat{\br}, \hat{h}]$, we have 
\footnote{Should two bands cross and $E_{m\bk}{\,=\,}E_{n\bk}$ at localized 
$\bk$ points, our implementation explicitly removes the divergence by setting 
$\br_{mn\bk}=0$ at such $\bk$.}
$\br_{mn\bk} = i\hbar \bm{v}_{mn\bk} / (E_{n\bk}-E_{m\bk})$. The matrix elements 
of the velocity can be approximated in the TB representation \eqref{eq:bloch} 
as \cite{Pedersen2001}
\begin{equation}
  \hbar \bm{v}_{mn\bk} \simeq \sum_{\alpha\beta}C^{m*}_{\alpha\bk} 
  C^{n}_{\beta\bk} \nabla_{\bk} 
  \bra{\chi_{\alpha\bk}} \hat{h} \ket{\chi_{\beta\bk}},
  \label{eq:vmn}
\end{equation}
where $\bra{\chi_{\alpha\bk}} \hat{h} \ket{\chi_{\beta\bk}}$ are the matrix 
elements of the TB Hamiltonian in the reduced Bloch representation. In a 
Slater-Koster framework, their $\bk$-dependence is explicitly known and, 
consequently, the $\bk$-derivative appearing in \Eqref{eq:vmn} can be directly 
computed once the TB Hamiltonian is specified.

%-------------------------------------------------------------------------------
\subsection{Self-energy approximation}
%-------------------------------------------------------------------------------

The COHSEX (Coulomb hole and screened exchange) approximation of Hedin 
\cite{Hedin1965} has been widely used to describe correlation effects in 
excited states \cite{Hybertsen1986,Onida2002,Bruneval2006}. It approximates the 
electronic self-energy as instantaneous in time and comprising two physical 
contributions: $\Sigma^\text{cohsex} = \Sigma^\text{coh} + \Sigma^\text{sex}$. 
The term
\begin{equation}
  \Sigma^{\text{sex}}(\br,\br', t) \equiv i G(\br t;\br't^+) \, 
    W(\br,\br';\omega=0)
  \label{eq:Sigma-sex}
\end{equation}
describes a statically screened exchange interaction, with $W(\br,\br';\omega)$ 
representing the dynamically screened Coulomb repulsion in the random-phase 
approximation, and 
\begin{multline}
  \Sigma^{\text{coh}}(\br,\br',t) \equiv \tfrac{1}{2}\delta(\br-\br') 
  \\
  \times \Bigl[ W(\br,\br';\omega=0) - w(\br,\br') 
  \Bigr],
  \label{eq:Sigma-coh}
\end{multline}
where $w(\br,\br') \equiv e^2/(4\pi\epsilon_0|\br-\br'|)$ is the bare Coulomb 
repulsion. The instantaneous approximation to $\Sigma$ is justified in 
the present context because the self-energy in \Eqref{eq:dGdt-2} is only 
operative in the presence of the external field. It therefore defines the 
strength of the electron-hole interaction but not the quasiparticle 
renormalization of the Kohn-Sham bandstructure. (While dynamical screening is 
crucial to obtain the correct quasiparticle renormalization \cite{Hybertsen1986, 
Onida2002}, in our formulation, that effect is already encoded in $\hat{h}$, by 
construction.) Furthermore, an instantaneous self-energy is in line with the 
current understanding, at the level of the BSE, that it correctly captures the 
excitonic spectrum of semiconductors\,---\,because of the excitons' relatively 
long time-scales in comparison with the dynamical charge oscillations involved 
in screening \cite{Rohlfing2000, Schone2002, Marini2003}. 

Recalling the reasoning above for the subtraction in \Eqref{eq:bloch} of 
the self-energy calculated at equilibrium, we see that the contribution 
$\Sigma^{\text{coh}}$, being time-independent in this approximation, does not 
contribute to the time evolution of the distribution function defined by 
\Eqref{eq:bloch}. We therefore need only to consider the screened exchange 
contribution \eqref{eq:Sigma-sex}. In this regard, note that the approximation 
in \Eqref{eq:Sigma-sex} for $\Sigma^{\text{sex}}$ consists in a static $GW$ 
approximation \cite{Onida2002}; hence, if the term $-\Sigma[\tilde{G}^{<}]$ were 
not included in \Eqref{eq:dGdt-2}, one would be doubly correcting the 
quasiparticle bandstructure renormalization.

Our approximation to the self-energy therefore reads
\begin{equation}
  \Sigma_{mn\bk}\bigl[G^{<}(t)\bigr] = i\sum_{jl\bk'} 
  W_{mn\bk,\,jl\bk'} \, G_{jl\bk'}^{<}(t)
  \label{eq:Sigma}
\end{equation}
in an explicit Bloch representation. The matrix elements of the screened 
exchange interaction are defined as
\begin{multline}
  W_{mn\bk,jl\bk'} \equiv \int \!\! d\br d\br' \, \psi_{m\bk}^*(\br) 
  \psi_{l\bk'}^*(\br') 
  \\ \times W(\br-\br') \psi_{j\bk'}(\br) \psi_{n\bk}(\br').
\end{multline}
In the TB basis introduced in \eqref{eq:bloch}, these matrix elements read 
explicitly  \cite{Pedersen2014,Ridolfi2018,native2D}$ ^{,}$ \footnote{To handle 
the $\bk$-diagonal cases where $\bk=\bk'$, we use the regularization scheme 
adopted in \Ref~\onlinecite{Ridolfi2018}}
\begin{equation}
  W_{mn\bk,jl\bk'} \equiv \sum_{\bG} 
  {[ I^{\bG}_{j\bk',m\bk } ]}^* \, I^{\bG}_{l\bk',n\bk} \; W(\bk-\bk'+\bG),
 \label{eq:W-matrix}
\end{equation}
where $W(\bq)$ represents the zero-frequency limit of the screened Coulomb 
potential,
\begin{equation}
  W(\bq) \equiv \biggl(\frac{e^2}{2\epsilon_{0}\epsilon_d A} \biggr)
  \frac{1}{|\bq|\,(1+\lambda_0|\bq|)}.
  \label{eq:w-rpa}
\end{equation}
This expression corresponds to the regime of small $\bq$ calculated for a 
strictly 2D electron gas embedded in three dimensions \cite{Keldysh1973, 
Cudazzo2011, Trolle2017}. $A$ is the area of the crystal, $\lambda_0$ its 2D 
polarizability \cite{Cudazzo2011, Chernikov2014, Rodin2014}, and $\epsilon_d$ 
captures the average dielectric constant of the environment. For us, in 
practice, $\epsilon_{d} = (\epsilon_1+\epsilon_2)/2$ to capture the effect of 
static, uniform screening due to the top ($\epsilon_1$) and bottom 
($\epsilon_2$) media surrounding the target 2D crystal. The parameters 
$\epsilon_d$ and $\lambda_0$ must be given to completely specify the screened 
interaction \eqref{eq:w-rpa}.

The Bloch coherence factors appearing in \Eqref{eq:W-matrix} 
are defined as
\begin{align}
  I^{\bG}_{m\bk,m'\bk'} 
    & \equiv \bra{m\bk} e^{i(\bk-\bk'-\bG)\cdot\hat{\br}} \ket{m'\bk'}
  \\
    & = \int d\br\, \psi^*_{m\bk}(\br) e^{i(\bk-\bk'-\bG)\cdot\hat{\br}} 
    \psi_{m'\bk'}(\br)
    \nonumber
\end{align}
Expanding for the Bloch functions introduced in \Eqref{eq:bloch} 
yields\footnote{To fix the relative phases in the numerical diagonalization of 
the Bloch Hamiltonian for different $\bk$, we require $\sum_{\alpha}C_{\alpha 
\bk}^{n}$ to be real, as suggested in \Ref~\onlinecite{Rohlfing2000} and 
previously implemented in \Ref~\onlinecite{Ridolfi2018}.}
\begin{multline}
  I^{\bG}_{m\bk,m'\bk'} \simeq
  \sum_{\alpha}[\,C_{\alpha\bk}^{m} e^{i \theta_{\alpha \bk} }\,]^{*} \, 
  [\,C_{\alpha\bk'}^{m'} e^{i \theta_{\alpha \bk'} }\,] 
  \\ 
  \times e^{i(\bk-\bk'-\bG) \cdot \bt_\alpha}.
\end{multline}
The vectors $\bG$ in the above expressions belong to the reciprocal lattice.
However, we emphasize that they do not reflect any attempt to include 
local-field corrections, which would not be warranted in our Slater-Koster TB 
scheme. The sum over $\bG$ is needed in \Eqref{eq:W-matrix} to restore the 
symmetry in the interaction between an electron with crystal momentum $\bk$ and 
another with $\bk'$: Since $\bk,\bk'$ are restricted to the first BZ\,---\,but 
the Fourier components of the Coulomb interaction are not\,---\,the interaction 
should include not only $\bk$ and $\bk'$ but all the equivalent pairs of crystal 
momenta. The decay of $W(\bq)$ justifies retaining only $\bG=0$ in most cases. 
However, in 2D materials the exciton binding energies can be extremely 
large\,---\,this implies tightly bound excitons in real space and, consequently, 
slowly decaying wavefunctions in reciprocal space. Consequently, the 
wavefunctions of an exciton at the $K$ valley and another at $K'$ can overlap, 
which might lead to a non-negligible Coulomb matrix element between those 
states. In such cases, the absence of the equivalent valleys beyond the first 
BZ amounts to an artificial symmetry breaking in the system\,---\,the summation 
over $\bG$ ensures such symmetry is retained. In practice, we keep only the 
fewest $\bG$ necessary to obtain converged results. (While in the calculations 
for MoS$_2$ we found that including only $\bG=0$ is sufficient, additional 
vectors were found to be important in the case of hBN, and we will return to 
this point later.)

Finally, we point out that in the proposal to implement this scheme fully 
\emph{ab initio} \cite{Attaccalite2011}\,---\,where $\hat{h}$ represents the 
Kohn-Sham Hamiltonian obtained as a first step within DFT\,---\,the Hartree 
energy must be updated at every time step as well for consistency. In our 
formulation, the Hartree term does not play a role under the time evolution. 
This is motivated by the fact that, in a BSE approach to the two-particle 
problem at $T=0$, the Hartree term in the Hamiltonian \eqref{eq:H} generates the 
so-called ``exchange'' interaction in the BSE \cite{Rohlfing2000}, whose matrix 
elements have the form
\begin{equation}
  \int \!\! d\br \,
    \psi_{c\bk}^*(\br) \psi_{v\bk}(\br) w(\br-\br') 
    \psi_{v\bk'}^*(\br') \psi_{c\bk'}(\br'),
\end{equation}
where the labels $c$/$v$ stand for indices that run over the conduction/valence
bands only. When written in Fourier space, and neglecting local-field 
corrections, this expression reduces to $w(\bq{\,=\,}0) \delta_{cv} 
\delta_{c'v'}$. But the Kronecker delta $\delta_{vc}=0$ because a conduction 
band can never coincide with a valence one. Hence, the Hartree contribution 
drops from the BSE, which means it is irrelevant for excitonic effects (this is 
also the reason why, even when one does take local-field effects into account, 
the Hartree contribution tends to be much smaller than that arising from the 
screened exchange interaction). By definition, an orthogonal TB expansion of the 
Bloch states such as \Eqref{eq:bloch} ignores local-field corrections. 
Therefore, in our formulation the Hartree contribution remains implicitly as 
part of $\hat{h}$ without dynamical updates.

We wish to emphasize two important aspects of this strategy to tackle the 
interaction effects. The first is that, by formulating the problem in the form 
of \Eqref{eq:dGdt-2} where the self-energy dictates all interaction 
and/or relaxation effects, a multitude of extensions to the current 
approximations is straightforward\,---\,it requires only the specification of 
additional terms in the self-energy, but not an overhaul of the implementation. 
Hence, this formulation is intrinsically versatile and adaptable. (The most 
interesting extensions would arguably be to include the influence of coupling to 
other degrees of freedom, such as phonons, or specific models of disorder as 
physical sources of broadening.) The second aspect relates to the specific 
approximation for our self-energy in \Eqref{eq:Sigma}: 
\Ref~\onlinecite{Attaccalite2011} shows that, in linear order on the external 
field, this approximation is equivalent to a combined $G_0W_0$+BSE approach, 
which is the state-of-the-art combination to reliably describe excitonic effects 
in semiconductors \cite{Rohlfing2000, Onida2002}. It is therefore the most 
promising basis to describe interaction effects in optical response beyond 
linear order.

% ------------------------------------------------------------------------------
% FIGURE BEGINS
% ------------------------------------------------------------------------------
\begin{figure*}
\centering
\includegraphics[width=0.8\textwidth]{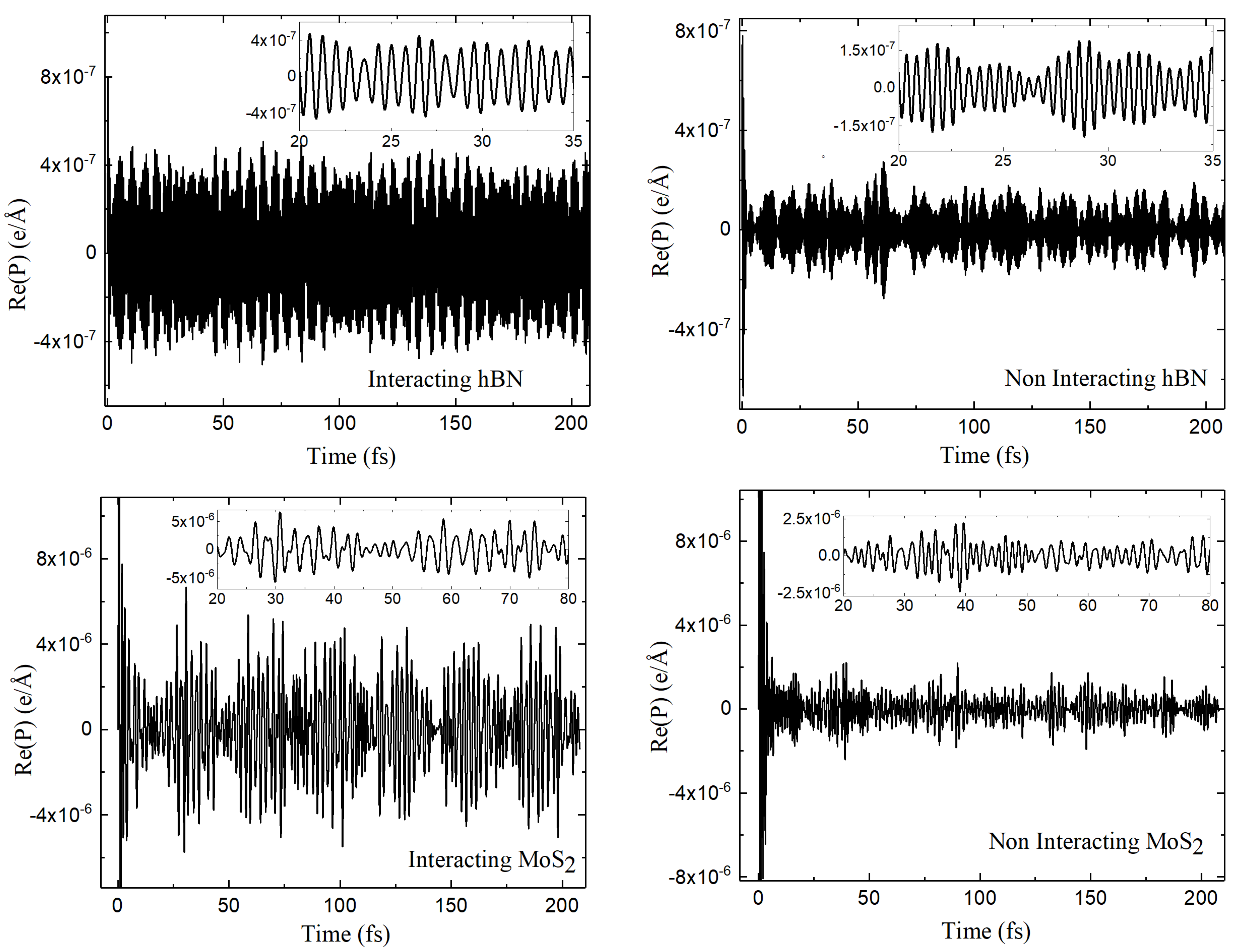}
\caption{
Time-dependent polarization calculated in response to a sub-fs optical pulse
for a monolayer of hBN (top row) and MoS$_2$ (bottom). For comparison, we show 
results with (left) and without (right) the effect of Coulomb interactions. 
[For hBN: $t_p=0.15$\,fs, $N_c=N_v=1$, $N_k^2=36^2$, $\tau=0.0125$\,fs, 
$E_0=0.1$\,mV/\AA. For MoS$_2$: $t_p=0.3$\,fs, $N_c=6$, $N_v=2$, $N_k^2=36^2$, 
$\tau=0.05$\,fs, $E_0=0.1$\,mV/\AA.]
}
\label{fig:polarizations}
\end{figure*}
% ------------------------------------------------------------------------------

%-------------------------------------------------------------------------------
\subsection{Fourier analysis of the optical response}
\label{sec:fourier}
%-------------------------------------------------------------------------------

To characterize the response in the frequency domain, we compute the discrete 
Fourier transform (FT) of the time-domain polarizability and electric fields. We 
define the discrete FT of a time-dependent signal $f(t)$ that is sampled at 
every constant interval $\tau$ as
\begin{equation}
  F_{\omega_k} \equiv \frac{1}{L} \sum_{n=0}^{L-1} f(t_n) e^{i\omega_k t_n}, 
  \label{eq:dft}
\end{equation}
where $t_n \equiv n\tau$, $L$ is the total number of time samples, $T \equiv 
L\tau$ is the total duration of the signal, and $\omega_k \equiv 2\pi k/T$ for 
$k\in\{0,1,...L-1\}$. It is obvious that the maximum frequency resolution of 
this procedure is $2\pi/T$ and, consequently, the total duration of the signal 
should in principle satisfy $T\gtrsim 2\pi/\gamma$ so that its Fourier spectrum 
has at least the resolution imposed by the characteristic broadening of the 
system, $\gamma$ [cf. \Eqref{eq:broadening} below]\,---\,this is one of the 
two compromises that ultimately determine the duration of the calculations in 
practice.

The other compromise is the time step, $\Delta t$, chosen to numerically 
integrate the equation of motion. Suppose, for example, that we wish 
to compute the response to a sinusoidal light field with a typical frequency of 
$\hbar\omega_0 \sim 1$\,eV (242\,THz). It should be clear that, because of the 
oscillatory nature of the solution for $P(t)$, one must set $\Delta t$ to a 
fraction of the fundamental period of the driving field to avoid accumulating 
numerical errors. If, for definiteness, one assumes 10 integration steps per 
fundamental period, we have $\Delta t = 2\pi/(10 \omega_0) \sim 0.4$\,fs. 
If we now seek a frequency resolution of, say, $10$\,meV, we must also have 
$2\pi \hbar/T \lesssim 10$\,meV, or $T \gtrsim 1000 \Delta t$. This means that 
the right-hand side of \eqref{eq:dGdt-2} must be evaluated at least 1000 times 
for such reasonable requirements. Of course, to ensure resolution of higher 
harmonics up to order $n$ of the fundamental frequency, one must replace 
$\omega_0\to n\omega_0$ in these estimates, whereby the numerical effort is seen 
to increase by a factor of $n$. 

The choice of the time step has also the fundamental constraint imposed by the 
Nyquist-Shannon theorem: Since the $\Delta t$ used in the numerical integration 
defines the smallest possible sampling interval ($\min \tau = \Delta t$), the 
theorem imposes the maximum frequency captured in a discrete FT to be $\pi/\tau$ 
and, consequently, one must ensure $\Delta t \le \tau < \pi / 
\omega_{\text{max}}$. As the energy scales of interest typically span several 
eV, $\Delta t$ must typically be well within the sub-fs range 
($\hbar\pi/1\,\text{fs} \simeq 2.1$\,eV) to allow a clean Fourier analysis 
(without aliasing, for example). 

These compromises can make the numerical integration time-consuming (see also 
\Sref{sec:notes} below). In contrast, the computation of the FTs appears 
``instantaneous'' when compared with the total time spent integrating the 
polarization up to $t=T$. For this reason, our discrete FTs have been computed 
by sampling with $\tau=\Delta t$ to maximize the amount of information, and do 
not require any optimization beyond the prescription in \Eqref{eq:dft}. 

Another advantage of computing the discrete FT as prescribed above using all the 
natural time steps from the numerical integration of \Eqref{eq:dGdt-2} is that 
we can avoid the phenomenon of frequency leaking and ensure we always obtain an 
exact representation of the continuous FT of the signal $f(t)$ whenever it 
consists of a series of discrete frequencies (like in a monochromatic wave). In 
order to see this, we recall a simple result from Fourier analysis and signal 
processing. Let the continuous FT of a signal $f(t)$ be defined by
\begin{equation}
  F(\omega) \equiv 
    \frac{1}{2\pi}\int_{-\infty}^{+\infty} \!\! f(t)\,e^{i\omega t}dt.
\end{equation}
$F(\omega)$ is related to $F_{\omega_k}$ defined in \Eqref{eq:dft} through
\begin{equation}
  F_{\omega_k} = \frac{2\pi}{T}\sum_{n=-\infty}^{+\infty} 
    \Bigl[ F(\omega-n\omega_s)     * \tilde{\delta}_{T/2}(\omega) 
    \Bigr]_{\omega=\omega_k},
  \label{eq:dft-vs-ft}
\end{equation}
where $*$ represents the convolution operation, $\omega_s \equiv 2\pi/\tau$ is 
the angular sampling rate, and 
\begin{equation}
  \tilde{\delta}_{T/2}(\omega) \equiv \frac{T e^{i\omega T/2}}{2\pi} 
    \text{sinc}\Bigl(\frac{\omega T}{2}\Bigr)
\end{equation}
is the FT of the rectangle function defined as unity for $0\le t\le T$ and zero 
otherwise. Now, if the signal is monochromatic with a frequency $\omega_0 < 
\omega_s$, we have $F(\omega)=F_0\,\delta(\omega-\omega_0)$ and, from 
\Eqref{eq:dft-vs-ft}, follows that
\begin{equation}
  F_{\omega_k} = \frac{2\pi}{T} F_0 \,\tilde{\delta}_{T/2}(\omega_k-\omega_0).
  \label{eq:Fwk}
\end{equation}
Noting that $\tilde{\delta}_{T/2}(0) = T/ (2\pi)$, the result \eqref{eq:Fwk} 
means that, by adapting the sampling rate $\omega_s$ (\ie, by chosing 
$\tau$) to $\omega_0$ (or vice-versa) such that $\omega_0$ is one of the 
frequencies $\{\omega_k\}$, we ensure that $F_{\omega_k=\omega_0} = F_0$ 
exactly and $F_{\omega_k}=0$ for all other discrete frequencies (\ie, the 
discrete FT recovers the \emph{exact} Fourier spectrum of the signal without any 
frequency leaking). This can be an important when extracting high-harmonic 
Fourier components from $P(t)$, which can be orders of magnitude smaller than 
the fundamental harmonic\,---\,we thus need to exclude or minimize all 
spurious effects, such as the inevitable frequency leaking that appears 
whenever $\omega_s$ is not matched to the relevant frequencies in the system's 
response signal. 

Fourier analysis of $P(t)$ will be used to obtain the nonlinear susceptibilities 
$\chi^{(n)}$ defined in \Eqref{eq:P-vs-chi-t}. In the frequency domain, that 
relation reads
\begin{multline}
  \epsilon_0^{-1} P(\omega) = 
  \sum_{n=1}^\infty \int\!\!d\omega_1 \cdots\! \int\!\!d\omega_n \,    
    \chi^{(n)}(\omega_1,\dots,\omega_n) 
    \\ 
    E(\omega_1)\cdots E(\omega_n) 
    \, \delta(\omega-\textstyle\sum_n\omega_n).
  \label{eq:P-vs-chi-w}
\end{multline}
By integrating \Eqref{eq:dGdt-2} in the presence of an external monochromatic 
field of frequency $\omega_0$, followed by the above-described approach to the 
discrete Fourier analysis, we extract all the $n$-th harmonic susceptibilities 
at once by computing
\begin{equation}
  \chi^{(n)}(\omega_0) = 
    \frac{P_{\omega_k=n\omega_0}}{(E_{\omega_k=\omega_0})^n \epsilon_0},
  \label{eq:chi-n}
\end{equation}
or the the associated optical conductivities: $\sigma^{(n)}(\omega_0) = 
-i\omega_0\epsilon_0 \chi^{(n)}(\omega_0)$.

%-------------------------------------------------------------------------------
\subsection{Relaxation and broadening}
%-------------------------------------------------------------------------------

The electronic system described by the equation of motion \eqref{eq:dGdt-2} 
accumulates all the energy transferred by the external radiation field (the 
self-energy is Hermitian). The corresponding runaway growth of the system's
polarization can cause numerical problems if \Eqref{eq:dGdt-2} needs to be 
integrated over a very large number of time steps.

A short-duration laser pulse is not numerically problematic in these conditions, 
as we shall see below. For the purposes of comparison with experiments 
using short-pulse excitation, one can incorporate a phenomenological energy 
broadening into the Fourier spectrum by a simple modification of \Eqref{eq:dft}:
\begin{equation}
  P_{\omega_k} = \frac{1}{L} \sum_{n=0}^{L-1} P(t_n) 
    e^{i (\omega_k +i\gamma)t_n}.
  \label{eq:dft-broad}
\end{equation}
In this way, broadening is introduced at the post-processing stage, where 
$\hbar\gamma$ defines the desired energy resolution (in our calculations we 
explicitly set it to the half-width at half-maximum of the lowest exciton peak 
that appears in the spectra used as reference to benchmark our results).

A scenario of continuous excitation without damping causes the amplitude of 
$\bm{P}(t)$ in the system to grow in time with a linear envelope, which 
introduces artifacts as $\omega\to 0$ in the numerical FT. In these cases, we 
found it desirable to introduce a phenomenological relaxation mechanism 
directly into the equation of motion \eqref{eq:dGdt-2}, which is then modified 
to
\begin{align}
  i\hbar\frac{\partial}{\partial t}\bm{G}_{\bk}^{<}(t)
  = & \Bigl[ \bm{H}_{\bk}(t) + \bm{\Sigma}_{\bk}[G^{<}(t)]
  - \bm{\Sigma}_{\bk}[\tilde{G}^{<}], \,\bm{G}_{\bk}^{<}(t) \Bigr]
  \nonumber \\
  & -i \hbar\gamma \bigl(\bm{G}_{\bk}^<(t) - \tilde{\bm{G}}_{\bk}^<\bigr).
  \label{eq:broadening}
\end{align}
The new term on the right-hand side promotes the return of the distribution 
function to equilibrium; the broadening parameter $\gamma$ is set to the 
target energy resolution as described above. 

Of course, both approaches are employed here as a phenomenological strategy to 
control the energy broadening of the final results, which is sufficient for the 
current purposes of this paper. But, as pointed out earlier, more sophisticated 
and microscopically motivated relaxation processes may be incorporated directly 
into the equation of motion as extensions of \Eqref{eq:broadening} 
\cite{Weissmer2018}.

%-------------------------------------------------------------------------------
\subsection{Implementation Notes}
\label{sec:notes}
%-------------------------------------------------------------------------------

\emph{Numerical integration}\,---\,After specifying the time dependence of the 
external field, we compute the time-dependent polarization according to 
\eqref{eq:P-def-G}. The set of coupled equations of motion represented 
by \Eqref{eq:dGdt-2} is integrated numerically using the second-order 
Runge-Kutta algorithm provided within the GSL library \cite{Galassi2009}. The 
choice of second-order is here a compromise between accuracy and expediency, 
since we wish to maintain the number of intermediate evaluations of the 
right-hand side of \eqref{eq:dGdt-2} as small as possible per 
time-step\footnote{Since the linear dimension of the problem is very large in 
virtually all cases of interest, there is no advantage in using higher-order 
methods which are only efficient when the function evaluations required to 
advance each time step are fast, which is not the case here.}. 

\emph{Problem dimension}\,---\,It is instructive to recall that the 
\emph{linear} dimension, $N_\text{tot}$ of the matrices in \Eqref{eq:dGdt-2} is 
defined by the total number of bands ($N_v$ valence and $N_c$ conduction) 
\emph{plus} the total number of $\bk$ points that need to be sampled in the BZ. 
Therefore, $N_\text{tot} = N_k^2 (N_c+N_v)^2$, where $N_k^2$ represents the 
total number of $\bk$ points ($N_k$ along each reciprocal direction). Since the 
Coulomb matrix elements \eqref{eq:W-matrix} are non-diagonal in $\bk$, their 
storage is the most costly since it requires $\sim N_\text{tot}^2/2$ entries in 
memory. Combining this with the fact that these matrices are to be multiplied 
several times for each time step of the Runge-Kutta integration leads to a 
numerical problem that quickly becomes challenging memory- and time-wise, 
even when considering minimal requirements of, say, $N_k=32$ and $N_c+N_v=2$.

\emph{Symmetry}\,---\,We will focus on 2D crystals with threefold 
symmetry, having point-symmetry group $D_{3h}$ or higher. This already covers 
the materials that are currently most actively studied, such as graphene and 
its derivatives, hBN, transition-metal dichalcogenides, silicene, stanene, 
germanene, and several others. This restriction is practical, not fundamental. 
It is adopted here because their linear and second-harmonic (SH)
susceptibility tensors are completely specified by computing only the diagonal 
component along $y$\cite{Boyd2008}. For our choice of lattice orientation in 
\Fref{fig:bands}(c), symmetry imposes $\chi_{xx}^{(1)}=\chi_{yy}^{(1)}$ and 
$-\chi_{yyy}^{(2)} = \chi_{xxy}^{(2)} = \chi_{xyx}^{(2)} = \chi_{yxx}^{(2)}$. 
From this point onwards we thus drop the Cartesian indices: $\bm{P}\to P(t) 
\equiv P_y(t)$, $\bm{E}\to E(t) \equiv E_y(t)$, and 
$\chi_{\lambda\alpha\cdots}^{(n)}\to\chi^{(n)} \equiv \chi^{(n)}_{yy\cdots}$. 

\emph{Nonlinear processes}\,---\,We will address only the $n$-th harmonic 
susceptibilities in this paper, which are a function of only one frequency. 
Hence, we will also adopt a simplified notation 
$\chi^{(n)}(\omega_1,\dots,\omega_n) \to \chi^{(n)}(\omega)$ throughout the 
remainder of this paper, where the single frequency $\omega$ is that of the 
driving field (\ie, the fundamental frequency when the field is monochromatic). 

\emph{Windowing}\,---\,We found that applying a window function to our numerical 
time series for $P(t)$ considerably reduces the effect of the transient 
background due to the field switch-on, which is an important consideration when 
resolving the nonlinear contributions (to be discussed below). Windowing 
consists in multiplying the original signal by a so-called window function, 
$\zeta(t)$, which is chosen to provide a desired redistribution of its Fourier 
spectrum. (This technique is frequently used to minimize frequency leaking 
in discrete Fourier analysis \cite{Orfanidis2010}.) The Fourier analysis 
described  in \Sref{sec:monochromatic-field} has been performed after applying a 
Hann window to the time-dependent polarization. In the notation of the previous 
section, this means replacing the time-dependent signal $f(t_n) \longrightarrow 
f(t_n)\,\zeta(t_n)$, where the Hann window function is defined as
\begin{equation}
  \zeta(t_n) \equiv 2\,\sin^2 \left( \frac{n\pi}{L-1} \right),
  \quad 0\le n \le L-1.
  \label{eq:Hann}
\end{equation}
In this formulation, the window function defines the envelope of the total time 
series $f(t_n)$. The Hann window does not introduce any additional parameter.

% ------------------------------------------------------------------------------
% FIGURE BEGINS
% ------------------------------------------------------------------------------
\begin{figure*}
\centering
\includegraphics[width=0.98\textwidth]{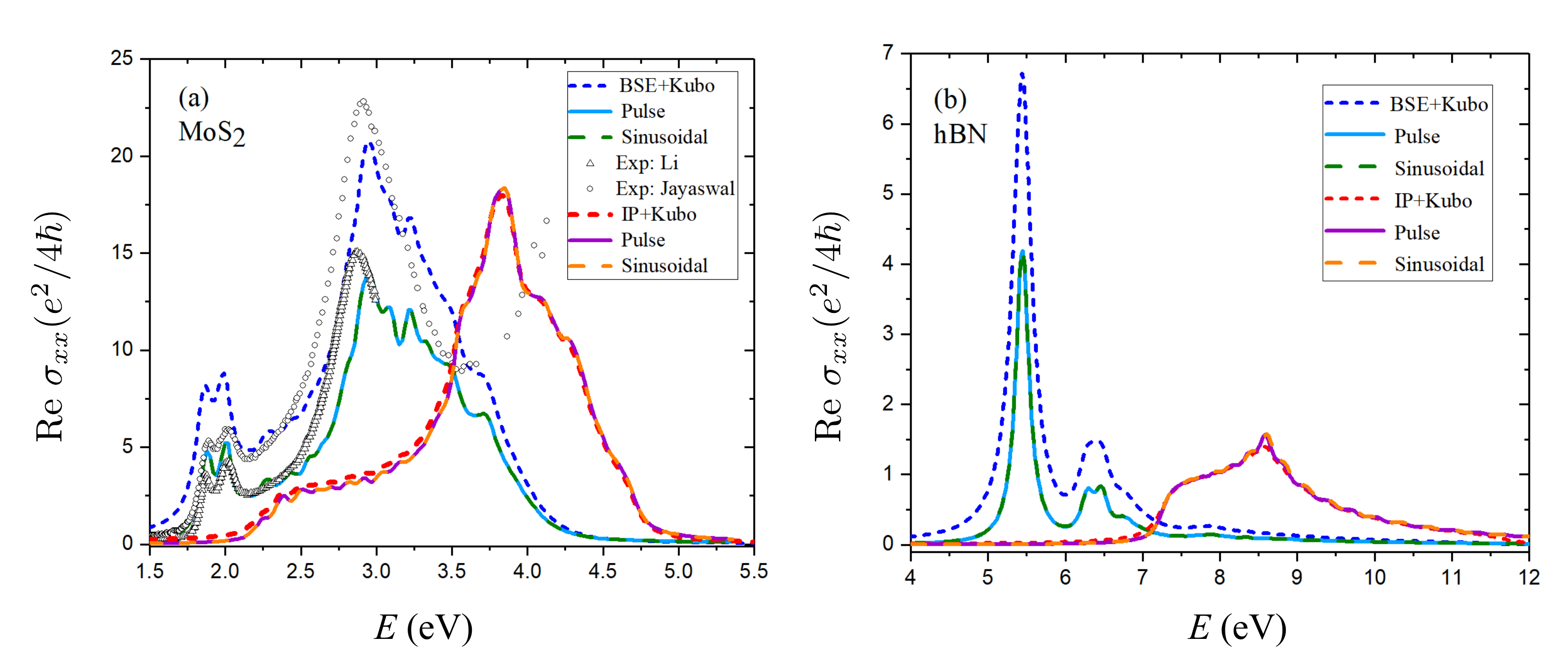}
\caption{
(a) Linear optical conductivity of MoS$_2$. The two curves represented by 
circles and triangles are experimental at room temperature \cite{Li2014, 
Jayaswal2018}. The lines labeled ``Pulse'' refer to the optical conductivity 
obtained from Fourier analysis of the time-domain polarizations shown in 
\Fsref{fig:polarizations}(c) and \ref{fig:polarizations}(d). The 
``BSE+Kubo'' (``IP+Kubo'') traces were obtained from the linear Kubo formula 
following the diagonalization of the Bethe-Salpeter (independent-particle 
Schr\"odinger, IP) equation for the same Hamiltonian. The ``Sinusoidal'' curves 
were calculated from the time-domain response to continuous monochromatic 
fields 
as described in the text. The broadening in our calculations was set to match 
that of the first exciton peak in the experimental traces shown in the figure.
(b) Linear optical conductivity of hBN with the same labeling convention as 
in (a). 
[For MoS$_2$: $t_p=0.3$\,fs, $N_c=6$, $N_v=2$, $N_k^2=36^2$, $\tau=0.05$\,fs, 
$E_0=0.1$\,mV/\AA, $\hbar\gamma = 0.05$\,eV. For hBN: $t_p=0.15$\,fs, 
$N_c=N_v=1$, $N_k^2=60^2$, $\tau=0.0125$\,fs, $E_0=0.1$\,mV/\AA, $\hbar\gamma = 
0.1$\,eV.]
}
\label{fig:sigmas}
\end{figure*}
% ------------------------------------------------------------------------------

%-------------------------------------------------------------------------------
\section{Illustration for \texorpdfstring{M\lowercase{o}S$_2$}{MoS2} and 
\texorpdfstring{\lowercase{h}BN}{hBN}}
\label{sec:results}
%-------------------------------------------------------------------------------

We illustrate the potential of this approach with the cases of MoS$_2$ and hBN, 
which have been chosen for different specific reasons. MoS$_2$ is the most 
widely studied representative of the family of 2D TMD semiconductors. It is now 
known that, except in the energy region of the bound A/B exciton series, the 
accurate description of the optical excitations across this family of compounds 
requires consideration of at least 6 conduction bands ($3\times 2$ for spin); 
this is due to the fact that the so-called C excitons involve contributions from 
the bands dispersing along $\Gamma$-K and $\Gamma$-M \cite{Qiu2013, Klots2014, 
Molina2015, Ridolfi2018} [cf. \Fref{fig:bands}(a)]. Since the spin-orbit-induced 
splitting of these bands is crucial for many of the unusual features in these 
materials, such as spin-valley locking \cite{Xiao2012, Mak2012, Zeng2012, 
WangReview2018}, a minimal model to describe the optical properties up to the 
energies of the C excitons requires in principle $2\times(3+1)=8$ bands to cover 
an energy span of $\sim 3$\,eV \cite{Pedersen2014, Ridolfi2018}. MoS$_2$ is then 
chosen as a representative of a system with a relatively demanding TB 
parameterization, and an example of how this approach can yield extremely good 
quantitative agreement with experiments and \emph{ab initio} calculations.
Our model for BN was deliberately selected to specifically analyze the opposite 
extreme of having only 2 bands in the problem; it will provide further insight 
into the role of the inter- and intra-band matrix elements of the dipole 
operator [cf. \Sref{sec:field}].

%-------------------------------------------------------------------------------
\subsection{Parameterization of Hamiltonians and interactions}
%-------------------------------------------------------------------------------

To describe the quasi-particle-corrected electronic structure of MoS$_2$ [\ie, 
the Hamiltonian $\hat{h}$ in \Eqref{eq:dGdt-2}] we consider the orthogonal 
Slater-Koster Hamiltonian proposed in \Ref~\onlinecite{Ridolfi2015}, which has 
been already demonstrated to capture extremely well the experimental optical 
absorption spectrum in a direct solution of the BSE \cite{Ridolfi2018}. It is 
built from an atomic basis comprising the three $p$ valence orbitals in each S 
and the five $d$ orbitals of Mo; since the spin-orbit coupling must 
necessarily be included to properly describe the splitting of the bands near the 
optical gap, the total dimension of the basis is then $N = 22$. Additional 
details of this TB model are described elsewhere \cite{Ridolfi2015}. The 
associated band structure is reproduced in \Fref{fig:bands}(a), reflecting the 
insulating ground state of a pristine monolayer with a direct gap at the $K/K'$ 
points. In order to directly compare our susceptibilities with experiments, a 
rigid blue-shift in the energies by $+0.07$\,eV has been incorporated in all the 
results shown below, in line with the procedure originally discussed in 
\Ref~\onlinecite{Ridolfi2018}.

For hBN, we resort to the orthogonal TB Hamiltonian parameterization proposed 
by Galvani \etal\cite{Galvani2016} as it provides a good description of the 
$GW$-corrected \emph{ab initio} band structure around the fundamental gap. 
It is a simple two-band (spin-degenerate) Hamiltonian which, in the notation 
introduced in \Eqref{eq:vmn}, reads
\begin{equation}
  \bra{\chi_{\alpha\bk}} \hat{h} \ket{\chi_{\beta\bk}}
  \mapsto
  \begin{bmatrix}
    E_b & -t\,\varphi(\bk)^* \\
    -t\,\varphi(\bk) & E_n 
  \end{bmatrix}\!\!.
  \label{eq:H-BN}
\end{equation}
where $E_b=3.625$\,eV represents the on-site energy at the boron atom, 
$E_n=-3.625$\,eV that at the nitrogen, $t=2.30$\,eV is the hopping integral, 
$\varphi(\bk) \equiv e^{i\bk\cdot\bm{\delta}_1} + e^{i\bk\cdot\bm{\delta}_2} + 
e^{i\bk\cdot\bm{\delta}_3}$ with vectors $\bm{\delta}_1 = 
\tfrac{a}{\sqrt{3}}(\tfrac{\sqrt{3}}{2}, -\tfrac{1}{2})$, $\bm{\delta}_2 = 
\tfrac{a}{\sqrt{3}}(0,1)$, $\bm{\delta}_3 = 
\tfrac{a}{\sqrt{3}}(-\tfrac{\sqrt{3}}{2}, -\tfrac{1}{2})$, and $a \simeq 
2.5$\,\AA{} is the hBN lattice constant. The associated band structure is 
reproduced in \Fref{fig:bands}(b). We note that this parameterization 
is accurate in the vicinity of the fundamental gap but does not faithfully 
reflect the actual dispersion over the entire BZ, especially near the $\Gamma$ 
point \cite{Arnaud2006, Galvani2016}. This limits the range of validity of our 
TB parameterization to particle-hole excitations with less than $\sim 
8\text{--}9$\,eV. 
In both materials, $\hat{h}$ is diagonalized in a uniform grid with ${N_k}^2$ 
points on the first BZ depicted in \Fref{fig:bands}(c).

For the purposes of benchmarking our calculations, the screened Coulomb 
interaction \eqref{eq:w-rpa} is parameterized in different scenarios for each 
material: For MoS$_2$ we chose the environment's dielectric constant as 
$\epsilon_d = 2.5$, which is appropriate for the air/silica interface 
($\epsilon_1 = 1$, $\epsilon_2 = 4$), and set the polarizability parameter $r_0 
= 13.55$\,\AA, which is know to produce good agreement with the measured exciton 
binding energies \cite{Zhang2014, Ridolfi2018}.
In the case of hBN, we used $r_0 = 10$\,\AA{} as suggested from DFT results 
\cite{Galvani2016}, while $\epsilon_d = 1$ so that we can directly compare our 
results with existing calculations for a free-standing hBN monolayer in vacuum. 
Finally, while for MoS$_2$ we found that considering only $\bG=0$ in the 
expressions for the Coulomb matrix element \eqref{eq:W-matrix} is sufficient, 
the case of hBN required the inclusion of at least 16 reciprocal vectors to 
recover the correct symmetry and degeneracy of the lowest excitonic states at 
the $K$ and $K'$ points.

%-------------------------------------------------------------------------------
\subsection{Excitons in the linear and nonlinear optical response of 
\texorpdfstring{M\lowercase{o}S$_2$}{MoS2} and 
\texorpdfstring{\lowercase{h}BN}{hBN}} 
%-------------------------------------------------------------------------------

Prior to discussing our results, we briefly overview the general features and 
calculations of the impact of electronic interactions (excitons) in the optical 
response of our two target materials, focusing especially on the nonlinear 
response.
MoS$_2$ has conduction and valence band extrema located at the two nonequivalent 
$K$ points of the hexagonal BZ. While the gap is indirect in multilayers, it 
is direct at those $K$ points for the monolayer \cite{Mak2010, Zhang2013}. A 
strong spin-orbit coupling splits the valence bands near $K$, generating two 
families of bound excitonic states traditionally labeled A and B \cite{Li2014, 
Zhu2015Sci, Qiu2013, WangReview2018}, and contributed primarily by the metal $d$ 
orbitals. The two A and B excitons ($E_A<E_B$) generate strong absorption 
peaks that define the optical absorption threshold at $E_A = 1.8 \pm 0.1$\,eV 
for a monolayer on silica, as can be seen in \Fref{fig:sigmas}(a) below. At 
higher energies, the absorption spectrum is dominated by the so-called 
C and D broad resonances that involve significant contributions from the 
chalcogen orbitals \cite{Qiu2013, Klots2014, Molina2015, Ridolfi2018}. 

Despite numerous studies of its linear optical response, there have been 
few experimental reports of the second and higher harmonic susceptibilities of 
MoS$_2$ over extended energy ranges. Examples are \Refs~\onlinecite{Malard2013} 
and \onlinecite{Kumar2013} that report the SH emission of both monolayer and 
trilayer in the region of the C resonance, and 
\Ref~\onlinecite{Pedersen2015trilayer} which reports similar measurements over 
the range $0.9\text{--}1.6$\,eV, but in multilayers.
SH calculations that include excitonic effects have been performed \emph{ab 
initio} in \Ref~\onlinecite{Attaccalite2014} and by Pedersen 
\etal\cite{Pedersen2014}, who used a perturbative formulation based on the 
solution of the BSE with a parameterized model analogous to ours. SH 
susceptibilities calculated with further simplified effective band models, 
applicable only to the region of the A and B peaks, have also been recently 
reported \cite{Glazov2017, Soh2018}.

In relation to hBN, the $p_z$ orbitals on B and N define the highest valence and 
lowest conduction bands which are separated by a large direct gap at the $K/K'$ 
points [see \Fref{fig:bands}(b)]. Experimentally, the optical gap is seen at 
$5.8$\,eV in bulk hBN \cite{Mamy1981, Watanabe2004} while values spanning 
$5.6\text{--}6.0$\,eV have been reported in optical absorption measurements for 
monolayers on quartz \cite{Song2010, Kim2012, WuPark2015}. Various excitonic 
characteristics and their impact in the optical response have been studied by 
DFT+$GW$+BSE \emph{ab initio} methods \cite{Wirtz2006, Arnaud2006, Arnaud2006b, 
Marini2008, Attaccalite2011, Attaccalite2011b, Yan2012a, Attaccalite2014, 
Cudazzo2016, Galvani2016, Koskelo2017, Ferreira2018, Attaccalite2018hBN}; 
\Ref~\onlinecite{Galvani2016} has provided, in addition, a real-space Wannier 
approximation to reduce the BSE to an effective exciton TB model. Pursuing a 
two-band model similar to that in \Eqref{eq:H-BN} and a screened interaction, 
Pedersen has solved the BSE equation and computed the SH susceptibility using an 
equilibrium second-order perturbative framework, and emphasized the importance 
of intra-band matrix elements of the dipole operator \cite{Sipe1995} in a 
length-gauge formulation of the coupling to the light field.

% ------------------------------------------------------------------------------
% FIGURE BEGINS
% ------------------------------------------------------------------------------
\begin{figure*}
\centering
\includegraphics[width=0.95\textwidth]{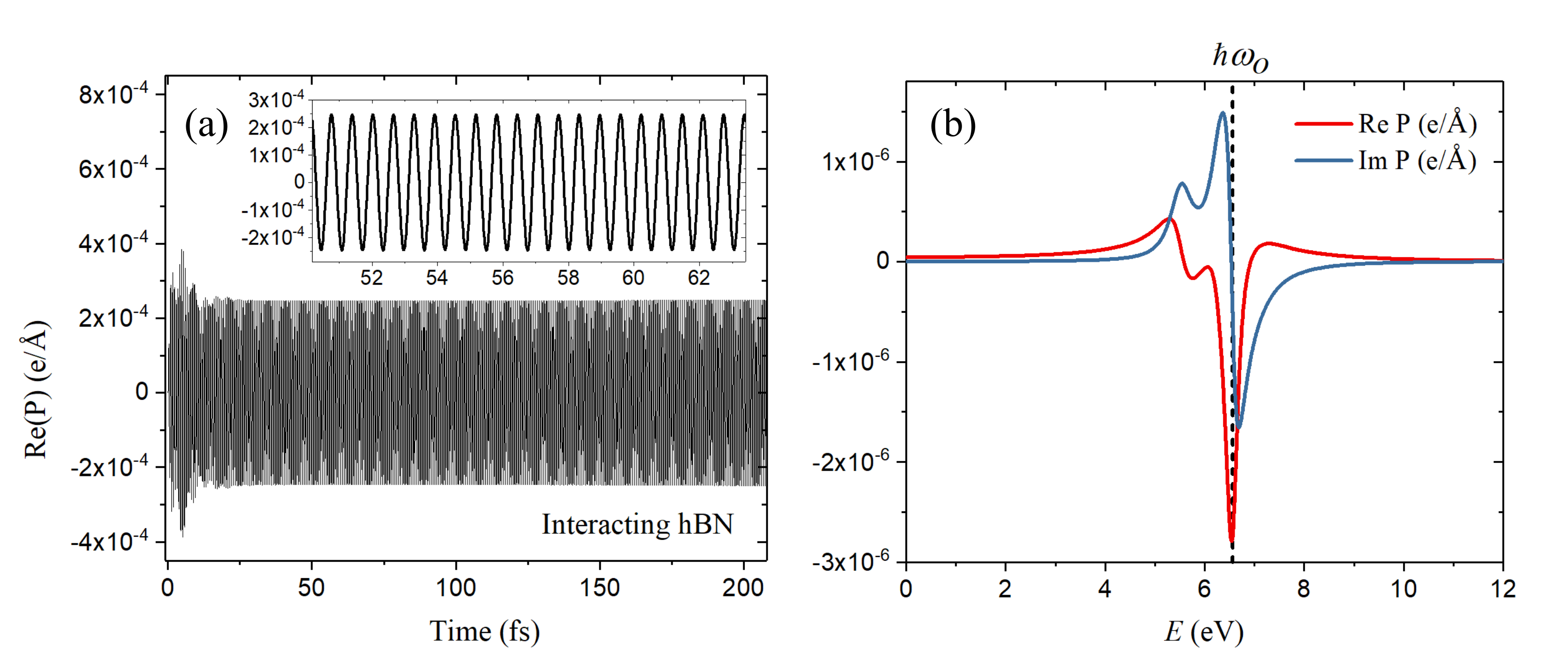}
\caption{
(a) Typical time-dependent polarization, $P(t)$, of a hBN monolayer in response 
to a continuous, monochromatic field with $\hbar\omega_0 = 6.65$\,eV. 
(b) The corresponding Fourier spectrum, $P(\omega)$. The dashed line marks 
the frequency of the external field. 
[$N^2_k=36^2$, $N_c=N_v=1$, $E_0=10^{-2}$\,V/\AA, $\tau=0.0125$\,fs, 
$\hbar\gamma = 0.1$\,eV] 
}
\label{fig:polarizations-sin}
\end{figure*}
% ------------------------------------------------------------------------------

%-------------------------------------------------------------------------------
\subsection{Linear response to an optical pulse}
%-------------------------------------------------------------------------------

The simplest perturbing field is that of a quasi-instantaneous pulse, which is 
particularly suitable to extract the linear response in a one-shot integration 
of \Eqref{eq:dGdt-2}. This is easily seen if one strictly sets $E(t) = E_0 
\delta(t)$ in \Eqref{eq:P-vs-chi-w} and ensures that $E_0$ is small; in lowest 
order 
\begin{equation}
  P(\omega) \simeq \frac{E_0}{2\pi\epsilon_0} \chi^{(1)}(\omega).
\end{equation}
Therefore, since an instantaneous pulse excites the system equally at all 
frequencies, the Fourier transform of $P(t)$ computed as the response to a 
single instantaneous pulse directly yields the linear susceptibility at all 
frequencies. We followed this approach to demonstrate that the results obtained 
by integrating the equation of motion \eqref{eq:dGdt-2} reproduce the linear 
susceptibility computed from direct diagonalization of the BSE combined with the 
Kubo formula. Note that, in general, once the linear response is established, 
the nonlinear susceptibilities are immediately defined as well since they are 
obtained from the same time-dependent polarizability of the system, as indicated 
in \Eqref{eq:chi-n}. This is particularly true with regards to the absolute 
magnitudes of the high-order susceptibilities because, by ensuring that the 
linear susceptibility is quantitatively accurate, we can subsequently rely on 
the predicted magnitude of the higher-order components.

To appreciate the details of the implementation, it is instructive to walk 
through some of its key aspects and intermediate results, which we will now do 
in relation to the response of a short-duration pulse. For the numerical 
implementation, we shaped the pulse as an inverted parabola,
\begin{equation}
  E(t) = E_0 (t_p-t) t/t_p^2, \quad 0\le t \le t_p,
  \label{eq:E-pulse}
\end{equation}
and $E(t) = 0$ beyond $t_p$. The amplitude was kept at $E_0=10^{-4}$\,V/\AA{} 
and we verified that nonlinear effects are absent in the range $E_0 
\sim 1\text{--}10$\,V/\AA, which is consistent with the expectation that, in 
general, nonlinear effects emerge when the field strength approaches the 
magnitude characteristic of atomic electric fields\cite{Boyd2008}: $E_\text{at} 
= e/(4\pi\epsilon_0a_0^2) \simeq 51$\,V/\AA{}. 

Figure \ref{fig:polarizations} shows the temporal profile of the induced 
polarization which has been integrated up to times in excess of 200\,fs after 
the initial excitation with the pulse \eqref{eq:E-pulse} ($t_p=0.15$\,fs for 
hBN; $t_p = 0.3$\,fs for MoS$_2$). The main panels show the total time series 
for $P(t)$ in both systems with and without the effect of the screened Coulomb 
self energy in the calculation. It is visible that $P(t)$ remains finite and 
undamped, reflecting the fact that this calculation did not include relaxation 
terms [\ie, $\gamma = 0$ in \Eqref{eq:broadening}].
Even without a detailed Fourier analysis, the time-domain picture reveals 
physically consistent signatures of the system's expected behavior: For 
example, we can identify by direct inspection an average period of $\sim 
0.48$\,fs for hBN and $\sim 1.0\text{--}1.4$\,fs for MoS$_2$ in the 
non-interacting polarizations [see insets of \Fsref{fig:polarizations}(b) and 
(d)]. These translate into characteristic energies of $\sim 8.6$\,eV and $\sim 
3\text{--}4$\,eV, respectively, which coincide with the energies at which the 
non-interacting absorption spectrum is maximal in each case (cf. 
\Fref{fig:sigmas} below). 

From a Fourier analysis of the $P(t)$ traces, we obtained the optical 
conductivities labeled ``pulse'' in \Fref{fig:sigmas}. The broadening was 
introduced as per \Eqref{eq:dft-broad} where we used the experimental width 
$\hbar\gamma \simeq 0.07$\,eV for MoS$_2$ and $\hbar\gamma = 0.1$\,eV for BN. 
The plot in \Fref{fig:sigmas}(a) pertains to MoS$_2$ and exhibits two types of 
comparison. The two traces represented by points were extracted from the 
experimental reports in \Refs~\onlinecite{Li2014} and \onlinecite{Jayaswal2018} 
for monolayers on SiO$_2$ substrates\,---\,one sees that our result reflected 
in the line labeled ``pulse'' describes all the experimental features very well, 
in particular the energies and spectral weight in the entire range of energies 
captured by our TB Hamiltonian ($\hbar\omega \lesssim 3.5$\,eV). The trace 
labeled ``perturbative'' has been computed by a direct application of the linear 
Kubo formula to the spectrum of the BSE (for exactly the same TB Hamiltonian and 
parameterized interaction used here) as described earlier in 
\Ref~\onlinecite{Ridolfi2018}. 
Finally, for reference and to further reinforce the substantial restructuring of 
the absorption spectrum brought about by the Coulomb interaction, the plot 
includes the conductivities obtained at the independent particle (IP) level  
(\ie, without the Coulomb self-energy). In quantitative terms, from the 
``impulse'' curves we extract a quasiparticle band gap of $2.18$\,eV and the 
lowest A/B excitons at $E_A=1.87$\,eV and $E_B=2.00$\,eV (binding energies 
$E^b_A=0.32$\,eV and $E^b_B=0.34$\,eV, respectively). This tallies well with 
results from angle-resolved photoemission spectroscopy \cite{Zhang2014}, as well 
as X-ray photoemission and scanning tunneling spectroscopy \cite{Chiu2015}, 
which place the band gap within $2.15\text{--}2.35$\,eV; this yields binding 
energies in the range $0.22\text{--}0.42$\,eV for the lowest A exciton.

The plots shown in \Fref{fig:sigmas}(b) reflect the corresponding calculations 
applied to the case of hBN. For reference, our explicit diagonalization of the 
BSE yields the lowest excitonic levels with the following energies and 
degeneracies (in eV): \{$5.44\,(\times 2)$, $5.96\,(\times 1)$, $6.27\,(\times 
2)$, $6.28\,(\times 1)$, $2.45\,(\times 2)$\}. As the quasiparticle band 
gap defined by our parameterization \eqref{eq:H-BN} is $7.25$\,eV, the binding 
energies of these excitonic levels are, respectively, \{$1.81$, $1.29$, $0.98$, 
$0.97$, $0.80$\}\,eV. These figures compare reasonably well with those
reported in \Refs~\onlinecite{Galvani2016} and \onlinecite{Arnaud2006b} using 
DFT+$GW$+BSE. Similarly to our result for MoS$_2$, the optical conductivity 
obtained for hBN with the time-domain framework reproduces the perturbative 
result, thereby validating our current implementation and the approximations 
involved. A characteristic of hBN is that the optical spectral weight is almost 
entirely concentrated at the exciton peaks, and is strongest at the lowest 
bright exciton. Indeed, the frequency dependence seen in \Fref{fig:sigmas}(b) 
reproduces extremely well the quantitative and qualitative features of the 
absorption spectrum obtained by several other groups \cite{Arnaud2006, 
Arnaud2006b, Wirtz2006, Marini2008, Yan2012a, Attaccalite2011, Pedersen2015hBN, 
Galvani2016, Ferreira2018}, and is compatible with the optical gaps of 
$5.6\text{--}6.0$\,eV reported experimentally \cite{Song2010, Kim2012, 
WuPark2015}. (Recall that we parameterized hBN in vacuum and, hence, both the 
$GW$ quasiparticle renormalization and the exciton binding would have to be 
adapted for a direct comparison with experiments.) Moreover, the absolute 
magnitude at the lowest exciton peak is here $\sigma^{(1)}(\omega = 5.4) \simeq 
4\,e^2/(4\hbar)$; this converts to an imaginary dielectric constant $\Im 
\varepsilon(\omega) = \Re \sigma^{(1)}(\omega) / (d \epsilon_0 \omega) \simeq 10 
$ (using $d = 3.3$\,\AA{} for effective thickness of a BN monolayer), which is 
entirely in line with the magnitude reported for this peak from first principles 
in \Refs \onlinecite{Arnaud2006, Ferreira2018}, as well as in optical absorption 
experiments with bulk BN \cite{Mamy1981}.

Common to both materials (and, by extension, to all 2D semiconducting materials) 
is the fact that, if one were to ignore the Coulomb interaction effects, the 
predicted absorption spectrum would be patently inaccurate: (i) it would appear 
with a large overall blue-shift [cf. non-interacting curves]; (ii) it would 
entirely miss the excitonic spectral weight that dominates near the absorption 
threshold. 

In the context of this paper, the most significant aspect of the results shown 
in \Fref{fig:sigmas} is that the time-domain calculation recovers the linear 
response function obtained perturbatively for the same microscopic 
parameterization of the system. We are thus in a position to explore the real 
power of the time-domain framework: its ability to naturally capture the 
response to arbitrary time-dependent fields, as well as to describe 
the nonlinear response in a rather expedite manner.

% ------------------------------------------------------------------------------
% FIGURE BEGINS
% ------------------------------------------------------------------------------
\begin{figure*}
\centering
\begin{minipage}{0.53\textwidth}
  \includegraphics[width=\textwidth]{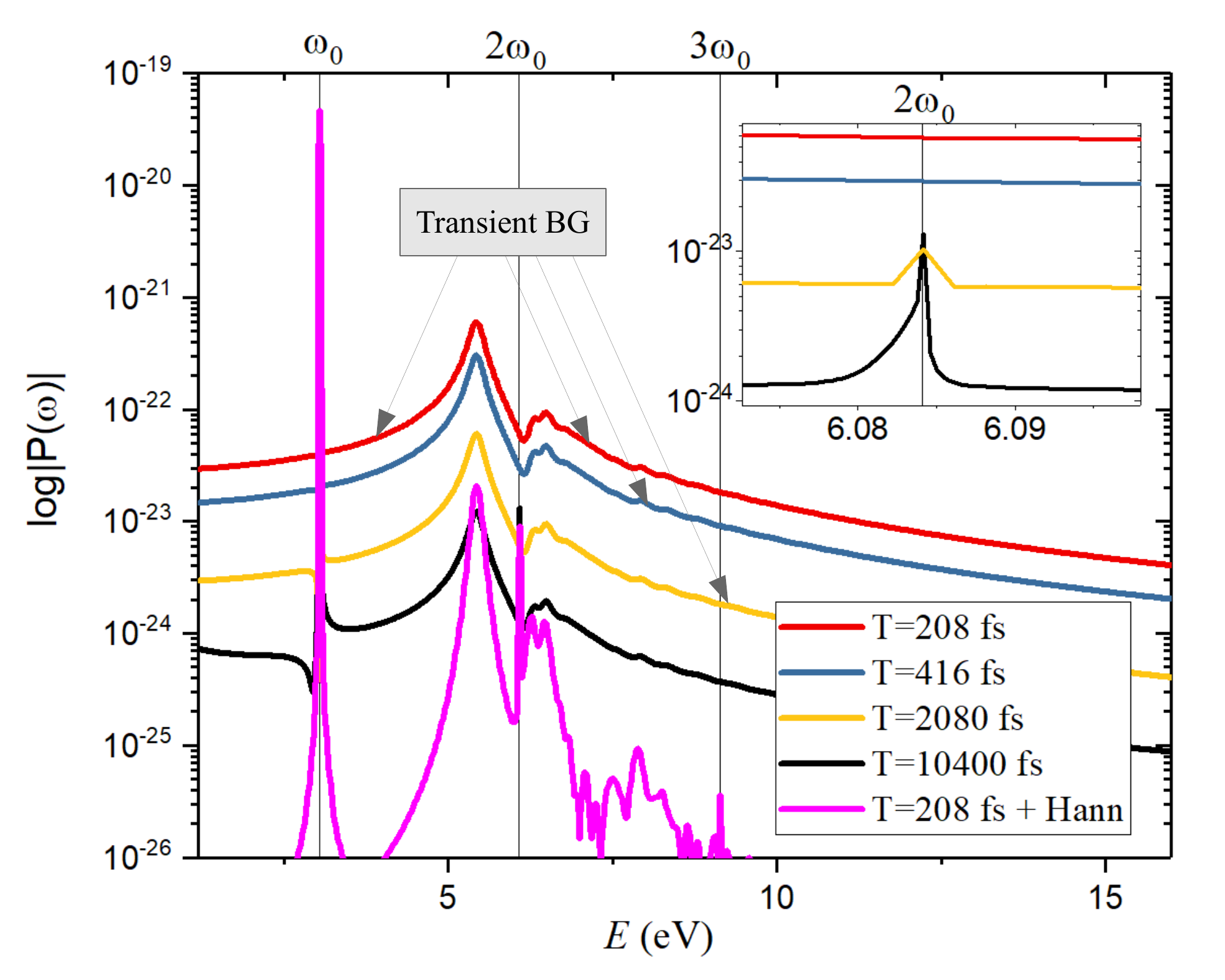}
\end{minipage}
\hfill
\begin{minipage}[c]{0.45\textwidth}
\caption{
Frequency spectrum of $|P(\omega)|$ for hBN as a function of the duration of 
excitation (total integration time, $T$) by a continuous, monochromatic wave 
($\hbar\omega_0 = 3.04$\,eV). The spikes at $n\omega_0$ define the amplitude of 
the $n$-th harmonic charge oscillations. Note how the peak at $2\omega_0$ 
remains occluded by the transient tail (follow the arrows labeled ``transient 
BG'') up to $T\sim 2080$\,fs (inset), and how the transient background is 
suppressed $\propto 1/T$ with increasing $T$. The four topmost curves reflect 
the raw frequency spectrum (without windowing); the bottom-most (magenta) 
illustrates the advantage of using a Hann window, which allows resolving up 
to the third harmonic with only $T=208$\,fs (without windowing, the trend from 
the other curves shows that $T$ would have to be ${\sim\,}10^{5}$\,fs 
to resolve the third harmonic). Note that the vertical scale is logarithmic.
[$N^2_k=36^2$, $N_c=N_v=1$, $E_0=0.01$\,V/\AA, $\tau=0.0125$\,fs, 
$\hbar\gamma = 0.1$\,eV]
\label{fig:P-vs-Tmax}
}
\end{minipage}
\medskip\hrule
\end{figure*}
% ------------------------------------------------------------------------------

%-------------------------------------------------------------------------------
\subsection{Nonlinear response to monochromatic fields}
\label{sec:monochromatic-field}
%-------------------------------------------------------------------------------

By definition, the high-harmonic susceptibilities $\chi^{(n)}(\omega_0)$ in 
\Eqref{eq:chi-n} represent the $n$-th order response of a system to a 
continuous, monochromatic wave at that frequency. More precisely, under a 
monochromatic perturbation of frequency $\omega_0$, the quantities 
$\chi^{(n)}(\omega_0)$ evaluated at the single frequency $\omega_0$ are 
sufficient to entirely specify the time or frequency dependence of the 
polarization. This offers a direct way to compute the high-harmonic 
susceptibilities by sending a light field
\begin{equation}
  E(t) = E_0 \sin(\omega_0 t),\qquad 0 \le t \le T,
  \label{eq:E-sin}
\end{equation}
computing $P(\omega)$, and repeating for as many frequencies $\omega_0$ as 
desired \cite{Attaccalite2013}. Of course, each calculation for a given 
frequency requires roughly the same duration as that for a quasi-instantaneous 
pulse which we described above. Therefore, the total time required to map 
$\chi^{(n)}(\omega)$ over a finite interval of frequencies will be comparatively 
much larger, in general, if a large number of frequencies is sought (by a factor 
that is roughly the number of such frequencies). Hence, despite the simplicity 
involved in extracting each $\chi^{(n)}(\omega_0)$ from a simple Fourier 
analysis as in \Eqref{eq:chi-n}, this strategy of sending one wave per frequency 
is the most time-consuming. A more expedite alternative is to excite the system 
with a pulse of finite duration (with enough bandwidth to span the range of 
frequencies of interest), followed by an order-by-order deconvolution of the 
field from the resulting $P(\omega)$, as determined by the relation 
\eqref{eq:P-vs-chi-w}. This route, however, relies on a much more involved 
post-processing and will not be pursued in the current paper. Given its 
simplicity, transparency and intuitive value, we shall instead proceed with the 
one-wave-per-frequency strategy to illustrate typical calculations.

Figure \ref{fig:polarizations-sin}(a) shows the calculated $P(t)$ for the hBN 
monolayer in response to a weak monochromatic field of the type \eqref{eq:E-sin} 
with $\hbar\omega_0 = 6.65 $\,eV ($\omega_0 = 9.97\times10^{15}$\,rad/s). A 
relaxation mechanism is now necessary to dissipate the energy that is constantly 
being pumped into the system by the continuous wave. We employed the scheme 
described by the second term in \Eqref{eq:broadening} with $\hbar\gamma = 
0.1$\,eV; other parameters are specified in the figure caption. 
As expected, $P(t)$ has now the temporal profile of a damped oscillator driven 
at a frequency $\omega_0$. Its Fourier analysis shown in 
\Fref{fig:polarizations-sin}(b) reveals a corresponding peak in $\Re P(\omega)$ 
at precisely $\omega_0$. By extracting $P(\omega=\omega_0)$ in this way for a 
number of distinct plane waves, we mapped the frequency-dependence of the linear 
susceptibility/conductivity and obtained the traces labeled ``sinusoidal'' in 
\Fref{fig:sigmas}. A direct inspection shows that they exactly follow the ones 
obtained with the pulse excitation described earlier.

There are important details worth emphasizing at this point in relation to the 
requirements for the total integration time, $T$. 
The \emph{first} consideration is that it clearly must be compatible with the 
desired energy resolution, say $\hbar\delta\omega$, which means that $T \gtrsim 
2\pi/\delta\omega$. 
The \emph{second} is that the system receives the incoming wave at $t=0$ on a 
state of equilibrium. Consequently, in addition to the \emph{driven} 
response there is also a \emph{transient} response to the sudden field turn-on 
that contributes to the polarization: $P(t) = P_\text{driven}(t) + 
P_\text{trans}(t)$ (precisely as in the classical driven oscillator where the 
solution of its equation of motion involves the sum of two such terms). With a 
damping rate $\gamma$, one expects the memory of the field turn-on to fade 
within a time $\sim 2\pi/\gamma$ and the corresponding decay of the transient 
component on the same time scale. In principle, one could discard the signal 
$P(t)$ up to that point in the Fourier analysis to minimize the transient 
effect\footnote{This has the same practical effect as, for example, 
adiabatically turning the field on over a time $\sim 2\pi/\gamma$, instead of a 
sudden turn-on.}. When combined with the energy resolution requirements, this 
roughly doubles the minimum value of $T$ up to which the equation of motion 
\eqref{eq:broadening} should be integrated. 

The \emph{third} consideration is that, in order to extract the nonlinear 
susceptibilities, one is interested in the frequency spectrum of the asymptotic 
component $P_\text{driven}(t)$, but not in that of $P_\text{trans}(t)$. The fact 
that the latter decays within a time $\sim 2\pi/\gamma$ is satisfactory only 
with regards to the linear response [meaning that, in practice, setting $T 
\gtrsim\sim 2\pi/\gamma$ is sufficient to guarantee an accurate result for 
$\chi^{(1)}(\omega)$ by following the procedure outlined in relation to 
\Eqref{eq:chi-n}]. However, the decay of $P_\text{trans}(t)$ might not be 
sufficient to resolve the nonlinear contributions to the polarization if the 
field amplitude ($E_0$) is small; in such a case, the transient contribution may 
conceal the higher harmonics. In order to illustrate this point explicitly, we 
plot in \Fref{fig:P-vs-Tmax} the absolute value of $P(\omega)$ obtained from 
$P(t)$ with different durations $T$. The case $T=208$\,fs corresponds to 
$2\pi\hbar/T \simeq 0.02$\,eV; according to the above, it should be adequate, in 
principle, to ensure an energy resolution of $0.1$\,eV in the derived response 
functions. 
However, we can see that, for the particular value of field amplitude $E_0$ 
used in \Fref{fig:P-vs-Tmax}, the SH peak at $2\omega_0$ is not resolved until 
$T \gtrsim 2000$\,fs, and the third harmonic remains entirely occluded by the 
transient background even for the longest durations of the 
excitation\footnote{For pedagogic purposes, we note that this is the same 
phenomenon that frequently happens experimentally, where one might need to 
acquire a week signal for times long enough to bring it above the noise floor. 
The fact that the background tails correspond to the transient component of the 
signal, as highlighted in \Fref{fig:P-vs-Tmax}, is confirmed by the fact that 
they scale $\propto 1/T$ as can be inspected in that figure.}. 
If this interplay between the total integration time, resolution, and field 
amplitude is not taken carefully into consideration, one risks 
erroneous results in the nonlinear susceptibilities. For example, 
suppose we were to blindly compute $\chi^{(2)}(\omega_0) = P(2\omega_0) / 
\epsilon_0 E(\omega_0)^2$, as prescribed by \Eqref{eq:chi-n}, directly from the 
red trace (labeled $T=208$\,fs) in \Fref{fig:P-vs-Tmax}: Rather than reflecting 
the actual SH susceptibility of the system, such result would correspond instead 
to the frequency spectrum of the transient background! In such case, however, 
an analysis of the field dependence would reveal that $\chi^{(2)}(\omega_0) 
\propto 1/E_0$, instead of being field-independent. This indicates that, 
ultimately, the computation of any $\chi^{(n)}(\omega)$ should be tested for 
field-independence within an adequate range of fields. (We ensured that to 
be the case in the results quoted in this paper for all susceptibilities.)

A \emph{fourth} consideration pertains to the more subtle fact that, 
rigorously, \Eqref{eq:P-vs-chi-w} is only applicable to non-resonant excitation, 
since it is a perturbative expansion in the external field. But when one is 
interested in mapping $\chi^{(n)}(\omega)$ for a given material, one is mostly 
looking at describing the resonant response\,---\,in the sense that $\omega$ 
(or $n\omega$) falls inside the spectrum of electron-hole excitations of the 
system. How, then, are we justified in using the expressions \eqref{eq:chi-n} 
if, under resonant excitation, the response of the system is not necessarily 
described by the series expansion \eqref{eq:P-vs-chi-w}? The answer lies in the 
finite broadening caused by the relaxation mechanism built into the time 
evolution [cf. \Eqref{eq:broadening}]: if $P(t)$ is integrated up to $T \gg 
2\pi/\gamma$ we gain enough energy resolution to appreciate that all states have 
an intrinsic lifetime and, therefore, the excitation is never strictly 
resonant\footnote{A loose way of illustrating the physical content of this 
statement is to think of a classical oscillator driven at the resonant 
frequency: Although its amplitude begins growing, upon waiting long enough one 
eventually sees that it does not diverge (strict resonance), reaching instead a 
finite maximum determined by the damping constant.}; in these conditions, the 
relation \eqref{eq:P-vs-chi-w} is justified and a valid means of obtaining the 
susceptibilities.

% ------------------------------------------------------------------------------
% FIGURE BEGINS
% ------------------------------------------------------------------------------
\begin{figure*}
\centering
\hspace*{-0.5cm}\includegraphics[width=1.06\textwidth]{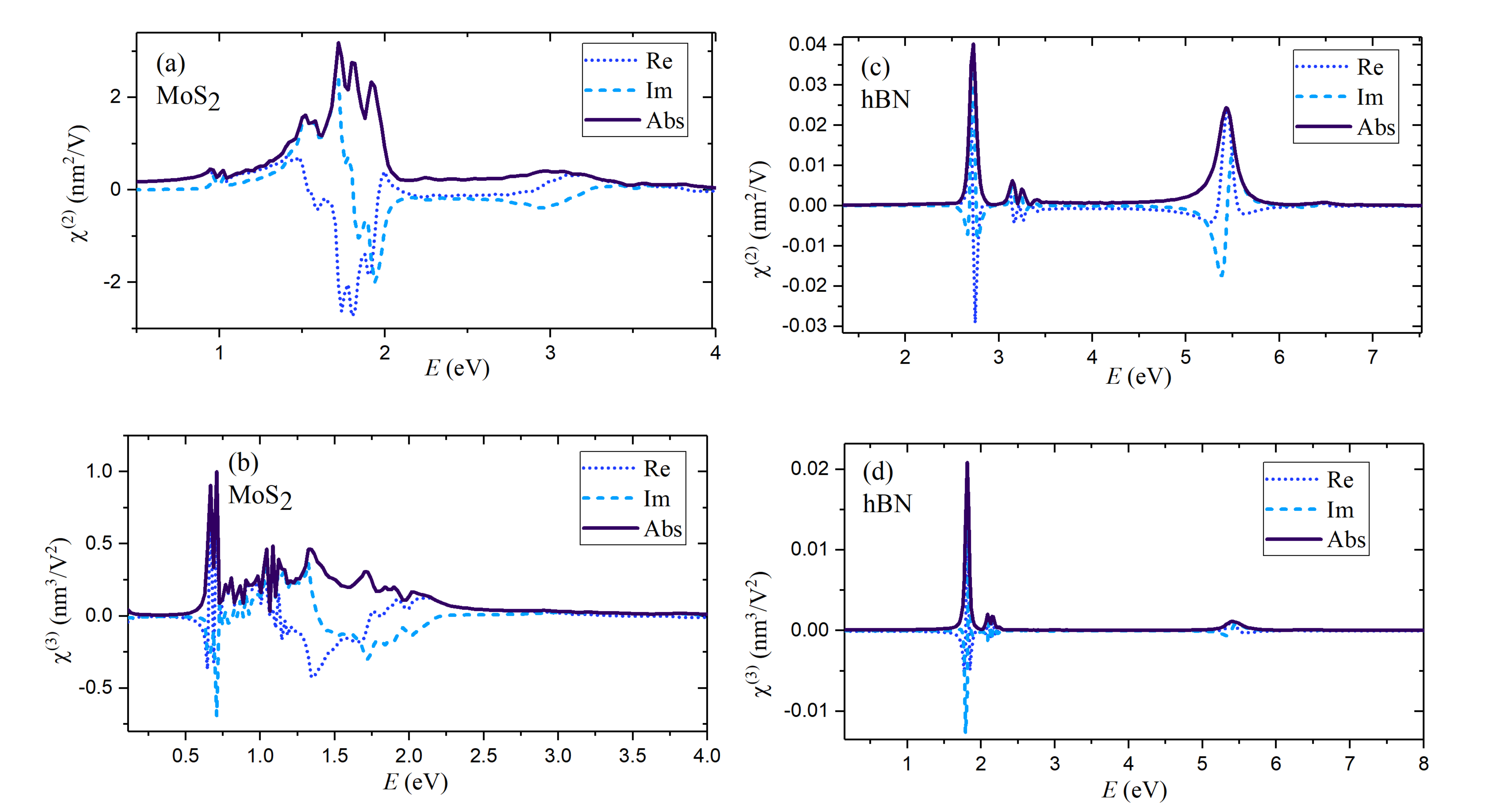}
\caption{
(a-b) Second- and third-harmonic susceptibilities of MoS$_2$. 
(d-e) Likewise, for hBN. Even though, we plot the susceptibilities in the 
extended frequency ranges shown in each panel for reference, recall that the 
underlying band structures are truncated; this limits the reliable ranges of 
validity of $\chi^{(2)}$ ($\chi^{(3)}$) to $\hbar\omega \lesssim 1.75 \, 
(1.17)$\,eV for MoS$_2$ and $\hbar\omega \lesssim 4.5 \,(3.0)$\,eV for hBN.
[For MoS$_2$: $N^2_k=36^2$, $N_c=6$, $N_v=2$, $E_0=0.01$\,V/\AA{}, 
$\tau=0.05$\,fs, $T=208$\,fs, $\hbar\gamma = 0.05$\,eV. For hBN: $N^2_k=60^2$, 
$N_c=N_v=1$, $E_0=5\times 10^{-4}$\,V/\AA{} $(\chi^{(2)})$, $E_0=1\times 
10^{-2}$\,V/\AA{} $(\chi^{(3)})$, $\tau=0.0125$\,fs, $T=208$\,fs, $\hbar\gamma = 
0.1$\,eV.] 
}
\label{fig:chi-2-3}
\end{figure*}
% ------------------------------------------------------------------------------

Having taken these aspects into account, we obtained the converged 
results shown in \Fref{fig:chi-2-3} for the second- and third-harmonic 
susceptibilities of MoS$_2$ and hBN. [For reference, we display the 
corresponding results without Coulomb interaction in \Fref{fig:chi-2-3-nonint}]. 
In each case, the features we obtained here for the SH susceptibility with 
explicit account of interactions compare well with other recent calculations. 
For MoS$_2$, we obtain SH magnitudes of 
$|\chi^{(2)}(\hbar\omega=0.9)| \simeq 0.4$\,nm$^2$/V (A/B exciton features) and 
$|\chi^{(2)}(\hbar\omega=1.5)| \simeq 1.5$\,nm$^2$/V (C exciton feature). The 
former compare well with the corresponding values $\simeq 0.12$ and $\simeq 
1.0$ obtained by Trolle \etal\cite{Pedersen2014} using a parameterized TB model 
to solve the BSE; the latter tally with the value $\simeq 0.7$\,nm$^2$/V 
($2.6\times 10^{-6}\,\text{esu}$) obtained \emph{ab initio} by Gr\"uning \etal{} 
\cite{Attaccalite2014-errata} at the C-exciton peak. 
Experimentally, Li \etal\cite{LiRao2013} reported 
$|\chi^{(2)}(\hbar\omega=1.53)| = 8.8 \times 10^{-31}$\,mC/V$^2$ $\simeq 
0.1$\,nm$^2$/V, while Woodward \etal{} \cite{Woodward2017} extracted 
$|\chi^{(2)}| = 0.02$\,nm$^2$/V and $|\chi^{(3)}| = 0.17$\,nm$^3$/V$^2$ from 
harmonic generation in MoS$_2$ under a laser field with $\omega_0 \simeq 
0.8$\,eV (1560\,nm). Our results in \Fref{fig:chi-2-3} for MoS$_2$ yield 
$|\chi^{(2)}(\hbar\omega=0.8)| \simeq 0.24$\,nm$^2$/V and 
$|\chi^{(3)}(\hbar\omega=0.8)| \simeq 0.25$\,nm$^3$/V$^2$. We consider them to 
be in reasonable agreement with experiments even though such comparisons are 
delicate because of a large dispersion in the magnitudes reported experimentally 
for high harmonic susceptibilities\footnote{Perhaps not surprisingly since these 
frequencies lie near resonant features of the susceptibilities and, as a result, 
the magnitude of the response varies noticeably under small deviations from 
those resonances. Small deviations can appear routinely due, for example, to 
sample variability, different pulse shape or duration, the use of different 
substrates, lack of temperature control, and thermal effects due to exposure 
to highly energetic laser beam}. 

In relation to hBN, the first remark is that, even though our TB model contains 
only 2 bands and we consider only inter-band matrix elements of the dipole 
operator, we capture a clearly finite $\chi^{(2)}(\omega)$ with all the 
excitonic features previously observed in \emph{ab initio} calculations 
\cite{Wirtz2006, Arnaud2006, Arnaud2006b, Marini2008, Attaccalite2011, 
Attaccalite2011b, Yan2012a, Attaccalite2014, Cudazzo2016, Galvani2016, 
Koskelo2017, Ferreira2018, Attaccalite2018hBN}, as well as in perturbative 
calculations based on parameterized TB models \cite{Pedersen2015hBN}. 
Our magnitudes, on the other hand, appear to be underestimated in comparison 
with these previous calculations by roughly one order of magnitude. For example, 
while the magnitude of our $\chi^{(2)}(\omega)$ in \Fref{fig:chi-2-3}(c) is 
$\simeq 0.04$\,nm$^2$/V at its strongest peak ($\hbar\omega \simeq 1.7$\,eV), 
the same peak has been reported with an intensity $\sim 0.2$\,nm$^2$/V, in both 
\emph{ab initio} \cite{Attaccalite2014-errata, Wang2015, Lucking2018} and 
parameterized \cite{Pedersen2015hBN} calculations. We address this discrepancy 
in the next section, but we point out that the only experimental report we are 
aware of quotes \cite{LiRao2013} $|\chi^{(2)}(\hbar\omega=1.53)| = 3 \times 
10^{-32}$\,mC/V$^2 \simeq 0.003$\,nm$^2$/V.

In relation to calculations accounting for excitonic effects beyond second 
order, we are only aware of those by Attaccalite \etal 
\cite{Attaccalite2018hBN} who calculate the frequency-dependent susceptibility 
for two-photon absorption in hBN. Unfortunately, that corresponds to the 
response function $\chi^{(3)}(\omega;\omega,\omega,-\omega)$ and not the 
third-harmonic $\chi^{(3)}(3\omega;\omega,\omega,\omega)$ that we can compute 
with our current implementation.

%-------------------------------------------------------------------------------
\section{Discussion}
\label{sec:discussion}
%-------------------------------------------------------------------------------

\emph{Trade-offs in the time domain}\,---\,%
Two key practical advantages of a time-domain formulation are: (i) 
\Eqref{eq:dGdt-2} condenses the problem in a formally simple expression which is 
suitable for a general-purpose implementation; (ii) it entirely circumvents 
the explicit calculation of each order of a perturbative expansion on the 
strength of the external field. Even though expressions have been given for some 
nonlinear susceptibilities, for both independent \cite{Armstrong1962, 
Sipe1993, Sipe1995} and interacting \cite{Leitsmann2005} electrons, the terms 
contributing to each order quickly proliferate and become cumbersome to handle 
already at the second order, especially when they incorporate excitons 
\cite{Leitsmann2005}. Besides, the actual calculation of each order inevitably 
demands a numerical integration over the BZ, even when using the simplest 
underlying Hamiltonians \cite{Pedersen2015hBN,Fabio2016}. Convergence of 
such integrations can become numerically challenging due to the presence of 
singularities in the spectral representation of the perturbative series that 
must be integrated.

On the other hand, the main trade-offs of a time-domain approach are: (i) the 
need of a post-processing stage that must be adapted to the information one 
desires to extract from the polarization (Fourier analysis, deconvolution, 
etc.); (ii) the total duration of $P(t)$ that must be acquired if one is 
interested in nonlinearities of very high order, or to achieve very high energy 
resolution in the final result. The issue here is that each time step in the 
numerical integration of \Eqref{eq:dGdt-2} is costly because the electronic 
self-energy is non-diagonal in crystal momentum $\bk$. In this regard, the 
strategy described in \Sref{sec:monochromatic-field} to obtain 
$\chi^{(n)}(\omega)$ is one of the worst case scenarios since it requires 
launching one wave for each frequency $\omega$ of interest. [The results in 
\Fsref{fig:sigmas} and \ref{fig:chi-2-3} required integrating the equation of 
motion \eqref{eq:broadening} hundreds of times, given the frequency interval and 
resolution we sought.] Yet, using parameterized Hamiltonians and interactions 
makes such computation entirely feasible without extreme computational 
resources, while a fully \emph{ab initio} implementation faces stringent 
computational challenges\footnote{For example, while the one-wave-per-frequency 
approach was employed in \Ref~\onlinecite{Attaccalite2011} to study the 
nonequilibrium band populations, it could only analyze a select (two) 
frequencies.}.

\emph{Ultrafast optical processes}\,---\,This is where the current framework 
has a distinctive advantage: as the temporal profile of the exciting field can 
be arbitrary, this approach is best suited for realistic simulations of 
properties that are intrinsic of the time domain. These include simulating the 
response to pulsed excitation, response to purposefully tailored light pulses, 
or pump-probe scenarios. More interesting is the potential to simulate 
electronic processes at ultrafast timescales (\eg, femtoseconds) while natively 
accounting for electronic interactions; this will be possible by adequate 
extensions of the electronic self-energy to capture specific mechanisms of 
electronic relaxation (\eg, electron-phonon and electron-electron collisions). 
Finally, since this framework is non-perturbative in the external field 
\emph{and} captures the evolution of the distribution function out of 
equilibrium, it is also naturally suited to characterize absorption saturation 
and other combined nonlinear+nonequilibrium effects of interest for 
applications.

% ------------------------------------------------------------------------------
% FIGURE BEGINS
% ------------------------------------------------------------------------------
\begin{figure}
\centering
\includegraphics[width=0.48\textwidth]{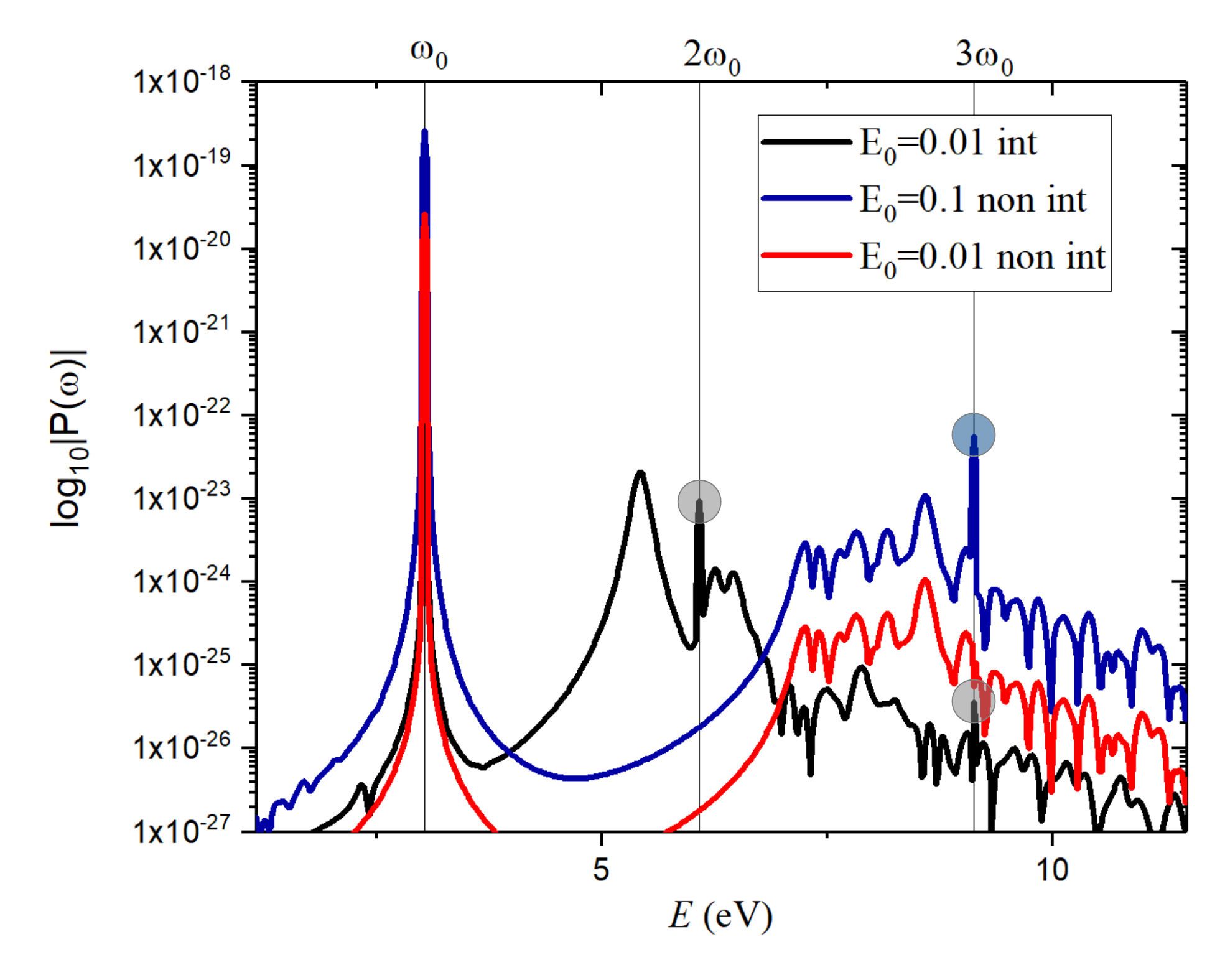}
\caption{
Frequency spectrum of $|P(\omega)|$ for hBN in response to a continuous 
monochromatic wave of frequency $\hbar\omega_0 = 3.04$\,eV, showing 
results obtained with (``int'') and without (``no int'') Coulomb interaction.
Note how, in the absence of interaction (blue and red curves), there is no 
second-harmonic response (no peak at $2\omega_0$), even when the field 
is strong enough to allow resolution of the third harmonic (blue curve). The 
shaded circles highlight the amplitude of the harmonic peaks at 
$\omega=n\omega_0$.
[$N^2_k=60^2$, $N_c=N_v=1$, $\tau=0.0125$\,fs, $T=208$\,fs, $\hbar\gamma 
= 0.1$\,eV] 
}
\label{fig:chi2-absent}
\end{figure}
% ------------------------------------------------------------------------------

\emph{Intra-band transitions}\,---\,As described in \Sref{sec:field}, we 
approximated the diagonal matrix elements of the dipole operator as $\br_{mm\bk} 
\simeq 0$, effectively assuming that only inter-band elements contribute to 
the optical response. Though this would be formally correct in linear response 
for a semiconductor at $T=0$\,K, the importance of the \emph{intra}-band 
contributions (IBCs) at higher orders has been a long and delicate subject of 
discussion (especially because it touches subtle aspects related to the 
choice of gauge and approximations to represent the coupling to the external 
electromagnetic field in the Hamiltonian \cite{Genkin1968, Sipe1995, 
Rzazewski2004, Fabio2017, Ventura2017, Taghizadeh2018}). A crucial disadvantage 
of this separation of contributions is that it breaks gauge invariance 
\cite{Foldi2017, Ernotte2018}.
The clearest and most striking example of the potential for inconsistencies 
arises in a two-band model: In the length-gauge and without interactions, 
\emph{equilibrium} second-order perturbation theory predicts the vanishing of 
the second-order susceptibility in the ground state of a semiconductor, 
irrespective of the underlying symmetry \cite{Sipe1995, Fabio2016}. The reason 
is trivial: the $n$-th order response involves the product of $n+1$ matrix 
elements defining a chain of transitions that must return to the initial 
state, which is impossible if $n=2$ and only inter-band transitions are 
included. Recently, it has been shown that adding excitons to the perturbative 
quadratic susceptibilities does not change the conclusion: IBCs are necessary to 
obtain a finite quadratic response in a two-band model \cite{Pedersen2015hBN}. 

In contrast, our calculations yield a finite $\chi^{(2)}$ [\Fsref{fig:P-vs-Tmax} 
and \ref{fig:chi-2-3}(c)] for a two-band description of hBN without IBCs; 
moreover, the obtained $\chi^{(2)}$ has the frequency-dependent features seen
in DFT+$GW$+BSE results \cite{Attaccalite2011, Pedersen2015hBN}. The fact we 
obtain a finite $\chi^{(2)}$ in a two-band description originates in the 
\emph{nonequilibrium} nature of our approach. To appreciate that explicitly, 
consider \Fref{fig:chi2-absent} where we show $P(\omega)$ under a monochromatic 
field with frequency $\omega_0$, calculated with and without interactions. In 
line with the simple argument given above, the non-interacting traces have no SH 
response [no peak at $P(2\omega_0)$], even when the field is strong enough to 
reveal a clear third-harmonic peak above the transient background [blue shaded 
disk; see also \Fref{fig:chi-2-3-nonint}(c)]. In contrast, the interacting trace 
does show a clear peak at $\omega{\,=\,}2\omega_0$ (gray shaded disk), the 
magnitude of which defines the $\chi^{(2)}(\omega)$ plotted in 
\Fref{fig:chi-2-3}(c) according to \Eqref{eq:chi-n}. 

The conclusion in \Ref~\onlinecite{Pedersen2015hBN} that IBCs are necessary to 
capture the SH susceptibility in a two-band model is \emph{conditioned} by 
their underlying assumption of quasi-equilibrium: the populations on each band 
remain unchanged by the external field. (In our formulation, this assumption 
amounts to not evolving the self-energy away from its equilibrium value, which 
implies setting $\bm{\Sigma}_{\bk}[G^<(t)] \to \bm{\Sigma}_{\bk}[G^<(0)]$.) 
But one can see from Eqs.~(4) and (5) of the cited reference that, by relaxing 
that assumption and explicitly integrating in time \emph{both} coherences and 
populations, one obtains additional contributions to the second-order response 
that involve only inter-band matrix elements. This is not surprising because 
one reason for the absence of purely inter-band contributions under the 
quasi-equilibrium assumption is the perfect Pauli blocking effect, which arises 
because the occupations $f_{m\bk}$ remain either 1 or 0 in that approximation, 
but not fractional. It thus follows that, when calculating the nonlinear 
response in a two-band model, one must not only include IBCs but also explicitly 
take into account the system's deviation from equilibrium. 

Therefore, strictly speaking, both our SH susceptibility and that in 
\Ref~\onlinecite{Pedersen2015hBN} for hBN are incomplete. As we highlighted 
above, although qualitatively identical, our calculated $\chi^{(2)}(\omega)$ is 
one order of magnitude below that reported in \Ref~\onlinecite{Pedersen2015hBN}; 
moreover, the latter tallies with the \emph{ab initio} result in 
\Ref~\onlinecite{Attaccalite2014} that employs yet another methodology. This 
can be an indication that, though not a negligible contribution, the deviation 
from equilibrium might be less dominant than the effect of IBCs. In our 
framework, a definitive conclusion requires restoring the intra-band matrix 
elements into the perturbation term \eqref{eq:U-field}, which is beyond the 
scope of the present paper.

\begin{figure}
\centering
\includegraphics[width=0.4\textwidth]{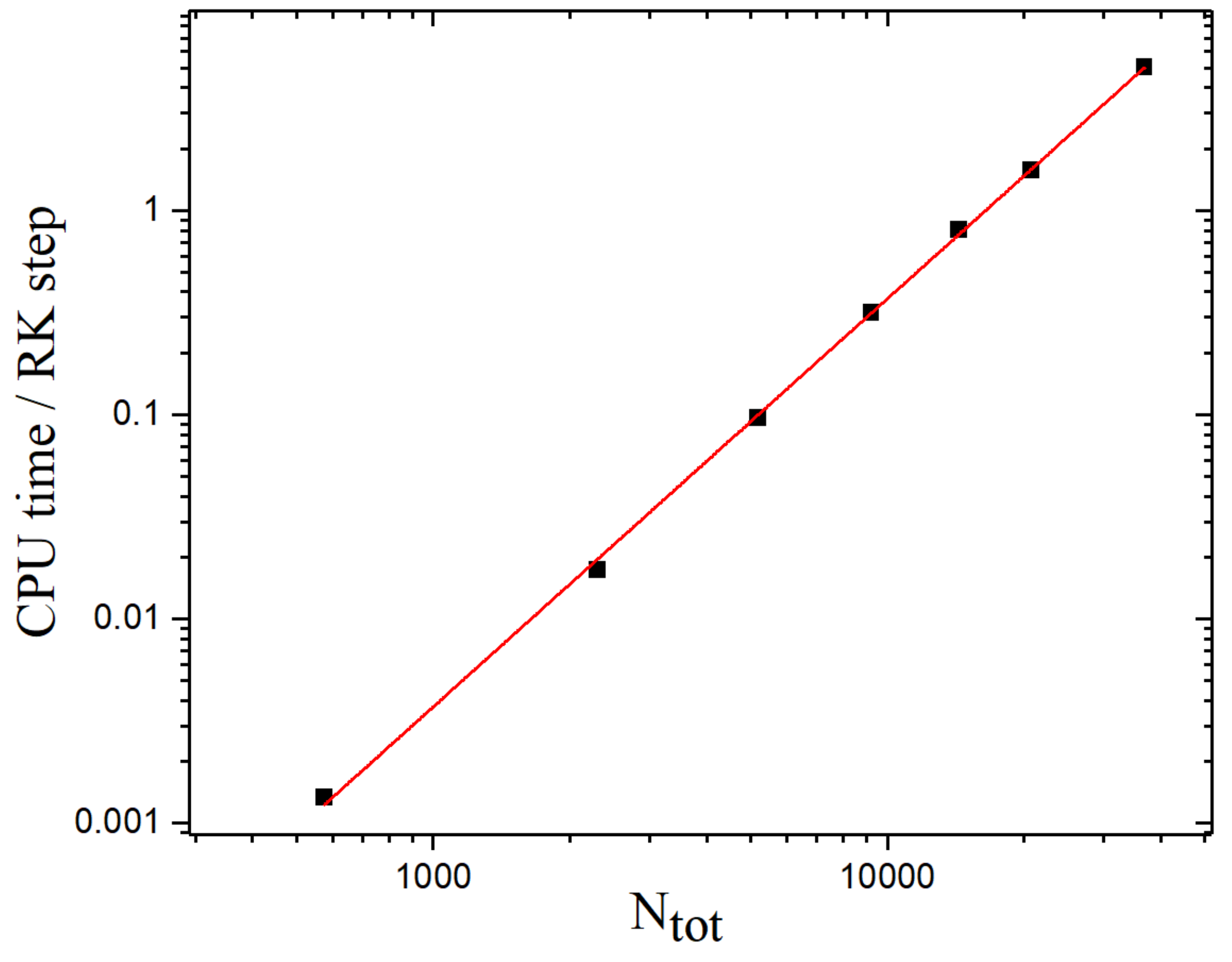}
\caption{
Typical scaling of the CPU seconds per integration time-step with the linear 
dimension of the Coulomb matrix, $N_\text{tot}=N_k^2(N_c+N_v)^2$. The line is 
$\propto N_{\text{tot}}^2$.
}
\label{fig:cpu}
\end{figure}

\emph{Numerical efficiency}\,---\,The numerical scaling of this framework is 
extremely simple and follows directly from the nature of the problem defined by 
the equation of motion \eqref{eq:dGdt-2}, the workhorse of the methodology. To 
integrate $P(t)$ in response to a \emph{single} external wave or pulse requires 
a total of $T/\Delta t = L$ time steps; when a Runge-Kutta algorithm is 
employed, one must recall that advancing one time step requires a number of 
intermediate evaluations that will be discarded, with more discarded the higher 
the order \cite{Abramowitz:1964}. (Our current implementation can indeed be 
sped up by a factor of two by switching to an integration rule that reuses all 
evaluations in subsequent steps.) 
As for storage requirements, it is desirable, for expediency, to store the 
Coulomb matrix elements \eqref{eq:W-matrix} which, being the only nondiagonal 
matrix in $\bk$, is what ultimately determines the storage needs. Since its 
linear dimension is $N_\text{tot}=N_k^2(N_c+N_v)^2$, it ultimately imposes a 
compromise between the number of points $N_k^2$ used to sample the BZ and the 
number of bands. But, as exemplified by our results in the 
one-wave-per-frequency (worst-case) scenario discussed in 
\Sref{sec:monochromatic-field}, we were able to include up to $N_c+N_v=8$ bands 
and $N_k^2\sim 36^2\text{--}40^2$ to describe MoS$_2$ still within reasonable 
computational resources. 
Finally, the calculation time per integration step scales $\propto 
N_{\text{tot}}^2$ because the evaluation of the right-hand side of 
\Eqref{eq:dGdt-2} can be coded as a matrix-vector product; this is explicitly 
shown in \Fref{fig:cpu}.

% ------------------------------------------------------------------------------
% FIGURE BEGINS
% ------------------------------------------------------------------------------
% ------------------------------------------------------------------------------

%-------------------------------------------------------------------------------
\section{Conclusion}
\label{sec:conclusion}
%-------------------------------------------------------------------------------

We have explicitly demonstrated that parameterized models are capable of 
retaining excellent agreement with experimental and \emph{ab-initio} optical 
spectra over large frequency ranges, while significantly alleviating the 
computational demands of the time-domain framework proposed by Attaccalite \etal 
\cite{Attaccalite2011} to study the response to arbitrary light fields. 
Therefore, our results broaden the practical reach of this general-purpose and 
versatile technique where multiple interacting and/or relaxation mechanisms can 
be incorporated in a systematic way, and which is natively suited to simulate 
the current frontier of ultrafast spectroscopy in solid-state materials. 
We have exposed in detail the relevant adaptations of the technique necessary 
for that, which will be of value to pursue further refinements and applications 
such as wave mixing or pump-probe simulations. 
Finally, we trust this will be a useful contribution to the current interest in 
robust general methods to tackle the combined nonlinear-nonequilibrium 
response of crystals under strong fields.

\bigskip

%-------------------------------------------------------------------------------
\begin{acknowledgments}
We acknowledge fruitful discussions with F. Hipolito, G. Ventura, M. D. Costa 
and J. C. Viana Gomes. 
Numerical computations were carried out at the HPC facilities of the NUS Centre 
for Advanced 2D Materials. We acknowledge the support and advice provided by 
M. D. Costa on various aspects of our numerical implementation and its 
optimization. 
This work was supported by the Singapore Ministry of Education Academic Research 
Fund Tier 2 grant number MOE2015-T2-2-059 (E. Ridolfi) and by the Singapore 
Ministry of Education Academic Research Fund Tier-1 internal grant reference 
R-144-000-386-114 (V. M. Pereira).
\end{acknowledgments}

\appendix

% ------------------------------------------------------------------------------
% FIGURE BEGINS
% ------------------------------------------------------------------------------
\begin{figure*}
\centering
\hspace*{-0.5cm}\includegraphics[width=1.05\textwidth]{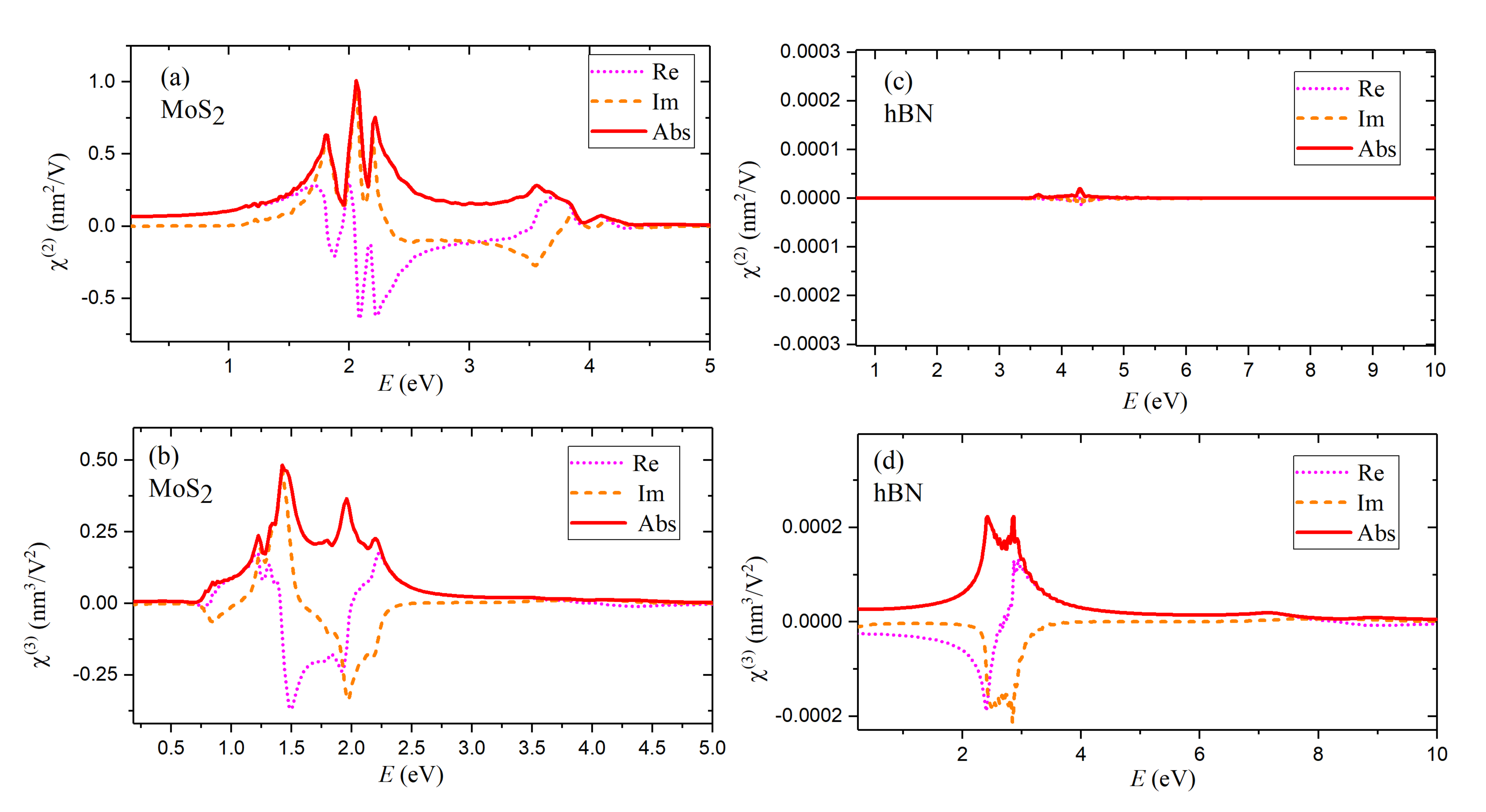}
\caption{
(a-b) Second- and third-harmonic susceptibilities of MoS$_2$ deliberately 
without Coulomb interactions. 
(d-e) Likewise, for hBN. Note how in (c) $\chi^{(2)}$ is within our noise floor 
(as expected for a non-interacting two-band model), in contrast with the 
interacting result shown in \Fref{fig:chi-2-3}(c).
[For MoS$_2$: $N^2_k=36^2$, $N_c=6$, $N_v=2$, $E_0=0.01$\,V/\AA{} 
$(\chi^{(2)})$, $E_0=0.05$\,V/\AA{} $(\chi^{(3)})$, $\tau=0.05$\,fs, 
$T=208$\,fs, $\hbar\gamma = 0.05$\,eV. For hBN $N^2_k=60^2$, $N_c=N_v=1$, 
$E_0=0.05$\,V/\AA{}, $\tau=0.0125$\,fs, $T=208$\,fs, $\hbar\gamma = 0.1$\,eV.] 
}
\label{fig:chi-2-3-nonint}
\end{figure*}
% ------------------------------------------------------------------------------

%-------------------------------------------------------------------------------
\bibliography{TMDs}
%-------------------------------------------------------------------------------

\end{document}